\documentstyle[12pt]{article}

\newcommand{\rf}[1]{(\ref{#1})}
\newcommand{\bea}{\begin{eqnarray}}
\newcommand{\eea}{\end{eqnarray}}
\newcommand{\g}{\gamma}
\renewcommand{\l}{\lambda}
\renewcommand{\b}{\beta}
\renewcommand{\a}{\alpha}
\newcommand{\n}{\nu}
\newcommand{\m}{\mu}
\newcommand{\th}{\theta}
\newcommand{\ep}{\varepsilon}
\newcommand{\vf}{\varphi}

\newcommand{\sg}{\sigma}
\newcommand{\dl}{\delta}
\newcommand{\prt}{\partial}
\newcommand{\dg}{\dagger}
\newcommand{\oh}{\frac{1}{2}}
\newcommand{\oq}{\frac{1}{4}}
\newcommand{\td}{\tilde}
\newcommand{\tr}{{\rm Tr}\,}
\newcommand{\Tr}{\tr}

\newcommand{\cL}{{\cal L}}

\newcommand{\ra}{\right\rangle}
\newcommand{\la}{\left\langle}

\newcommand{\slD}{D\!\! \!\! / \,}
\newcommand{\sld}{\partial \!\!\! /\,}
\newcommand{\Gf}{\Gamma_5}
\newcommand{\gf}{\g_5}
\newcommand{\tGf}{\td{\Gamma}_5}
\newcommand{\avGf}{\langle \tGf \rangle}

\newcommand{\CSn}{Chern-Simons number\ }

\def\void{}
\def\labelmark{}

\newenvironment{formula}[1]{\def\labelname{#1}
\ifx\void\labelname\def\junk{\begin{displaymath}}
\else\def\junk{\begin{equation}\label{\labelname}}\fi\junk}%
{\ifx\void\labelname\def\junk{\end{displaymath}}
\else\def\junk{\end{equation}}\fi\junk\labelmark\def\labelname{}}

{\ifx\void\labelname\def\junk{\end{array}\end{displaymath}}
\else\def\junk{\end{array}\right.\end{equation}}
\fi\junk\labelmark\def\labelname{}\def\junk{}
}

\newcommand{\beq}{\begin{formula}}
\newcommand{\eeq}{\end{formula}}
\newcommand{\beqv}{\begin{formula}{}}

\begin{document}
\topmargin 0pt
\oddsidemargin 5mm
\headheight 0pt
\topskip 0mm

\addtolength{\baselineskip}{0.20\baselineskip}
\rightline{NBI-HE-93-62}
\rightline{SWAT/14}

\hfill November 1993
\begin{center}

\vspace{36pt}
{\large \bf LEVEL CROSSING FOR HOT SPHALERONS}

\end{center}

\vspace{36pt}

\begin{center}
{\sl J. Ambj\o rn\footnote{The Niels Bohr Institute, Blegdamsvej 17,
DK-2100 Copenhagen \O , Denmark}
, K. Farakos\footnote{National Technical University of  Athens, Department of
Physics, GR 157 80, Athens, Greece.},
S. Hands\footnote{Department of Physics, University of Wales, Swansea,
Singleton Park, Swansea SA2 8PP, U.K.} ,
G. Koutsoumbas$^2$ and G. Thorleifsson$^1$}

\vspace{36pt}

\end{center}

\vfill

\begin{center}
{\bf Abstract}
\end{center}

\vspace{12pt}

\noindent
We study the spectrum of the Dirac Hamiltonian in the presence of
high temperature sphaleron-like fluctuations of the electroweak
gauge and Higgs fields, relevant for the conditions prevailing in the
early universe. The fluctuations are created by numerical
lattice simulations. It is shown that a change in Chern-Simons number
by one unit is accompanied by  eigenvalues crossing zero and a
change of sign of the generalized chirality $\tGf= (-1)^{2T+1} \gf$
which labels these modes. This provides further evidence that
the sphaleron-like configurations observed in lattice simulations
may be viewed as representing continuum configurations.

\vspace{24pt}

\vfill

\newpage

\section{Introduction}\label{sec-intro}

The baryonic current is anomalous in the electroweak theory \cite{thooft}.
This leads to a non-conservation of baryon- and lepton-number.
It is of little importance in the present day universe
where the processes which violate baryon- and lepton-number conservation
are exponentially suppressed. There is evidence that this suppression
is not present at the high temperatures prevailing in the
early universe \cite{aaps}, and it  has played an increasing role in
attempts to explain the baryon asymmetry observed in the universe today.

Unfortunately almost any interesting question related to
the high temperature phase of the electroweak theory
is difficult to address by analytical tools, due to the infrared
singularities of the associated high temperature perturbation theory.
A generic non-perturbative way to study the high temperature fluctuations
of the gauge and Higgs fields is by means of lattice simulations. If we
want to address questions related to the anomaly we encounter the
difficulty that topology is involved, an inherent continuum concept,
and it is important to verify that configurations generated on the lattice
during the  simulations will qualify as representatives of
continuum configurations.

In \cite{aaps} a real time evolution of classical gauge and Higgs fields
was used as an approximation to the high temperature fluctuations present
in the electroweak theory in the early universe. The baryon number violation
was studied by observing the change in the Chern-Simons number during the
time evolution. Since the temperature fluctuations are not small in these
simulations it is important to ask if these fluctuations might spoil
the picture of level crossing, which somehow is the justification of
viewing the change in \CSn as responsible for a change in fermion number.

In \cite{af} it was shown that the gauge field fluctuations which lead
to a change in Chern-Simons numbers at the same time cause the lowest
eigenvalue of the {\it massless} Dirac Hamiltonian to cross zero, in agreement
with the level crossing picture where such a crossing results in the
conversion of particles to anti-particles (or vice versa). In addition the
measurement of the spatial extension of energy lumps associated with
the fluctuations corroborated the interpretation of these as
representing continuum physics. However, in the full electroweak
theory an important ingredient is the Higgs field. It  couples
to the fermions and is responsible for particle masses in the
broken phase. At first sight it is maybe surprising that the level crossing
picture is still true in this case since the eigenstates of the full Dirac
Hamiltonian are massive when one includes the coupling to the Higgs field.
This is in particular true if we consider the ground state configuration in
the broken phase : $\phi = \phi_0$, $A_\mu = 0$. The situation is
different in the presence of a sphaleron. For a sphaleron configuration there
is precisely one normalizable eigenmode and it has zero
energy \cite{ringwald,kunz}. As a consequence a gauge-Higgs configuration
which changes continuously from one vacuum configuration to a
neighboring one, passing through a sphaleron configuration on the
way, will trigger an adiabatic change of the lowest positive eigenvalue
such that it passes through zero and ends up as the ``highest'' negative
eigenvalue \cite{kunz}.

In  many ways the situation around a sphaleron configuration is similar
to the situation around the simplest non-abelian monopole.
In that case there exists an index theorem for open three-dimensional
space \cite{calias}  which tells us that the difference  $n$ between
the number of  normalizable eigenfunctions of positive chirality and the
normalizable eigenfunctions of negative chirality, of
the Dirac Hamiltonian with eigenvalue zero,
is related to the charge of the monopole $Q$ by
\beq{*1}
Q= \frac{4\pi n}{e}.
\eeq
$n$ can be expressed in terms of the Higgs field:
\beq{*2}
n = \frac{1}{4\pi} \int_{S_\infty} d^2x_i \ep_{ijk} \ep^{abc}
\td{\phi}^a \prt_j \td{\phi}^b \prt_k \td{\phi}^c,
\eeq
where $\td{\phi}^a=\phi^a/|\phi|$ for a Higgs field in the adjoint
representation. Eq. \rf{*2} reflects that $|\phi|$ is assumed to be
different from zero at spatial infinity and consequently that
the map
$x\in S^2_\infty \to \td{\phi}(x) \in S^2$ has winding number $n$.
In the case of the sphaleron we do not have a similar topological
interpretation since the Higgs field is a complex doublet, i.e. it
has four real components and  we will have a map
$x\in S^2_\infty \to \td{\phi}(x) \in S^3$. This map is always
contractable to the trivial map ($\pi_2 (S^3) = 0$). However,
for a continuous family of static fields
$\phi (x;\tau), A_i(x;\tau) $, $\tau \in [0,1]$,
which interpolates between two neighboring vacua when $\tau$
changes from 0 to 1, there has to be at least
one value $\tau_0$ of $\tau$ such that the Higgs field
$\phi(x,\tau_0)$ has a zero. In the case of the electroweak theory
and the so-called minimal path which connects two neighboring
vacua the sphaleron is precisely such a configuration. If we
follow the minimal path we know that  at the sphaleron
there will be a normalizable eigenmode of the Dirac operator and
it has zero eigenvalue. It is intuitively reasonable that a zero
of the Higgs field is a necessary condition for having a zero
eigenvalue of the full Dirac operator. However, there is
to our knowledge no topological theorem which tells us that
the full Dirac operator has eigenvalue zero for some $\tau_0$.
This is in contrast to the case of the massless Dirac operator,
which only couples to the gauge field, and where
it follows from the Atiyah-Patodi-Singer index theorem that the
number of eigenvalues of the Dirac Hamiltonian which crosses zero
is directly related to the change in Chern-Simons number
\cite{cdg,christ,kiskis}.

Recently a number of articles have addressed the question of
level crossing for the full Dirac  Hamiltonian in the electroweak
theory \cite{kunz1,aj,amn}. The proof of the validity of the
level crossing picture has either been of numerical nature, for a specific
chosen path in  $(\phi, A_i)$ configuration space, or it has used
some special symmetry properties of the sphaleron solutions and is
only valid for some range of the coupling constants.
In this article we make an attempt to investigate the validity of the
level crossing picture in the case of fluctuations of the gauge {\it
and especially the Higgs fields} which are typical for the high 
temperatures of the early universe.

There is one more reason to focus on the Higgs field. The role of this
field in the high temperature fluctuations, and in particular its
relation to the masses of the fermions and therefore to the
level shifting, is not very clear. If the
symmetry is restored the classical expectation value of the Higgs
field is equal to zero, even if we have imposed complete gauge fixing.
However, the field still couples to the fermions and the statement that
$\la \vf \ra_T =0$ refers to the thermal average $\la \cdot \ra_T$. When we
discuss temperature induced transitions of gauge and Higgs fields between
neighboring vacua connected by large gauge transformations we
deal with dynamical processes caused by thermal fluctuations, and
it is strictly speaking misleading to refer to thermal averages. 
In fact we want in the
discussion of the anomaly and level shifting to
use an approximation where the gauge and Higgs fields are treated as
background fields {\it and even above the symmetry restoration temperature
we can talk about ``sphaleron-like'' transitions}. In the simulations
we perform the expectation value $\la |\vf|^2\ra$ will typically be
quite small compared to the classical vacuum expectation value 
$\la |\vf|^2 \ra _0$ at
zero temperature, due to temperature fluctuations. This indicates that the
system is in the symmetric phase rather than in the broken phase.
It is nevertheless perfectly possible (for a finite volume)
to talk about the fields as located in the neighborhood of a definite
vacuum and the transition rate between such vacua. 
The picture becomes especially
clear if one talks about the family of all fields which interpolate between
the neighborhoods of two different vacua, and then focuses on those
configurations in which the high energy thermal fluctuations have been
stripped off.
By such a  procedure it becomes clear that the low frequency modes of the
gauge fields can be viewed as interpolating between the different vacua
and that the \CSn will change by one unit during the interpolation 
\cite{af}.
In this article we will use the spectral flow of the Dirac operator as
a means to include the Higgs field in the analysis.

The rest of this article is organized as follows: In sec. \ref{sec-mod}
we define the ``reduced'' fermionic part of the electroweak model we will use.
In sect. \ref{sec-lat} we show how to
discretize the corresponding Dirac Hamiltonian
on a three-dimensional spatial lattice and especially how to define the
concept of chirality on this lattice. Sec. \ref{sec-cal} contains the 
numerical analysis of level crossing and chirality change for a 
number of ``hand made'' families of gauge-Higgs configurations. This 
analysis is made to compensate for the lack of mathematically rigorous 
results for Dirac Hamiltonians coupled to chiral fermions, results 
which we need when interpreting the level shifting and chirality
change in the numerical simulations.
In sect. \ref{sec-num} we describe the numerical
method used to simulate the gauge-Higgs part of the electroweak theory.
Sect. \ref{sec-mea} reports on
the measurements of spectral flow of the Dirac
operator for ``sphaleron-like'' transitions. Finally
sect. \ref{sec-dis} contains a
discussion of the results obtained.

\section{The model}\label{sec-mod}

Let us consider the electroweak theory where for simplicity we
ignore the hyper-charge sector.  It is of no importance for the
electroweak anomaly. We also restrict ourselves to
two  families. The generalization to the twelve existing
doublets is trivial. 

The fermionic part of the Lagrangian is given by:
\bea
\cL_f &=& \bar{\psi}_L \slD \psi_L + \bar{\chi}_R \sld \chi_R+
\bar{\psi'}_L \slD \psi'_L + \bar{\chi'}_R \sld \chi'_R
 \nonumber \\
&& - \bar{\psi}_L (h_u \td{\vf} \chi_{uR} +h_d \vf \chi_{dR})-
( h_u \bar{\chi}_{uR}{\td{\vf}}^\dg+h_d \bar{\chi}_{dR} \vf^\dg  )\psi_L
\nonumber \\
&& - \bar{\psi'}_L (h'_u \td{\vf} \chi'_{uR} +h'_d \vf \chi'_{dR})-
( h'_u \bar{\chi'}_{uR}{\td{\vf}}^\dg+h'_d \bar{\chi'}_{dR} \vf^\dg  )\psi'_L
\label{*3}
\eea
where the left handed doublets are
\beq{*3a}
\psi_L = \left( \begin{array}{c}  \psi_{u} \\ \psi_{d}
\end{array} \right)_L~~~~~~~~~~~
\psi'_L = \left( \begin{array}{c}  \psi'_{u} \\ \psi'_{d}
\end{array} \right)_L~~
\eeq
and the four right handed singlets are
$\chi_{uR},\chi_{dR},\chi'_{uR},\chi'_{dR}$.
With an abuse of notation we collect them as vectors, like the
doublets, even if they of course behave like singlets under electroweak
$SU(2)$ transformations:
\beq{*3b}
\chi_R = \left( \begin{array}{c}  \chi_{u} \\ \chi_{d}
\end{array} \right)_R~~
\chi'_R = \left( \begin{array}{c}  \chi'_{u} \\ \chi'_{d}
\end{array} \right)_R~~
\eeq
The Higgs doublet $\vf$ and its  conjugated field $\td{\vf}$
(which transforms as $\vf$ under electroweak transformations) are
defined as usual:
\bea
\vf &=& \pmatrix{ \phi_+ \cr \phi_0 \cr},\label{*4a} \\
\td{\vf}& =& \pmatrix{ ~\phi_0^* \cr -\phi_+^* \cr} = i\tau_2 \vf^*
\eea
There are four  independent Yukawa couplings $h_u$, $h_d$, $h'_u$ and $h'_d$.
As mentioned above we have only two independent couplings in the
corresponding lattice model and for simplicity we therefore
choose $h'_u=h_u$ and $h'_d = h_d$.
$h_u \approx h_d$ mimics the the quark sector (as the notation indicates),
while $h_d=0$ describes the  lepton sector.

It is well known that the chiral theory defined by \rf{*3}
can be given a formulation  as a vector-like theory of Dirac fermions
\cite{ls1}.
This is due to the fact that the representations of $SU(2)$
are real. We can therefore introduce the charge-conjugated fields
\beq{*3c}
\psi'^c_R = \ep C \psi'_L=
\left( \begin{array}{c}  \psi'^c_{dR} \\ -\psi'^c_{uR}\end{array} \right)
\eeq
where $\ep=i\tau_2$ acts on the isotopic indices and $C$ is the
charge-conjugation matrix, acting on spinor indices: $\psi^c \equiv C \psi$.
In a similar way we define
\beq{*3d}
\chi'^c_L =  C \chi'_R=
\left( \begin{array}{c}  \chi'^c_{dL} \\ -\chi'^c_{uL}\end{array} \right).
\eeq
If we use the chiral representation of the $\g$ matrices a Dirac fermion can
be written as
\beq{*3d1}
f_D =\left( \begin{array}{c}  f_L \\ f_R \end{array} \right).
\eeq
and we can now introduce the two Dirac spinors:
\beq{*Dirac}
\psi_D =\left( \begin{array}{c}  \psi_L \\ \psi'^c_R \end{array} \right),~~~~~
\chi_D = \left( \begin{array}{c}  \chi_{R} \\ \chi'^c_L \end{array} \right).
\eeq
This allows us finally to introduce the {\it eight}-component
spinor $\Psi$ by:
\beq{*eight}
\Psi = \left( \begin{array}{c}  \psi_D \\ \chi_D \end{array} \right),
\eeq
and using the matrix
\beq{*5}
M = \Phi \; \pmatrix{h_u & 0 \cr 0 & h_d\cr},~~~~~~~
\Phi = \pmatrix{ \phi_0^* & \phi_+ \cr -\phi^*_+ & \phi_0 \cr},
\eeq
we can write the Lagrangian \rf{*3} as follows
\beq{*6}
-\cL_f = \bar{\Psi} \pmatrix{ \slD &  -M \cr
                             -M^\dg & \sld} \Psi
\eeq
such that each entry of the matrix in eq. \rf{*6} is a
$2\times2$ matrix in isospinor indices and a $4\times4$ matrix in
spinor indices.

Note that \rf{*6} resembles the ordinary Dirac equation but
that {\it (i)} the mass  is spatially varying, depending on
$\vf$, {\it (ii)} the mass term may multiply the 1st and 2nd (and 3rd and
fourth) components differently if $h_u \neq h_d$ and, most importantly,
{\it (iii)} only the first 4 components ($\psi_D$) of the eight component
spinor $\Psi$
have gauge interactions. The last four ($\chi_D$) are gauge singlets.

The notation introduced  here follows \cite{aj}, which also introduced
the generalized chirality matrix $\tGf$ acting on the eight component
fermion field $\Psi$:
\beq{*8}
\tGf = \pmatrix{\g_5 & 0 \cr 0 & -\g_5}.
\eeq
The matrix $\tGf$ anticommutes with the matrix in \rf{*6} and also with
the matrix
\beq{*8a}
\Gamma_0 = \pmatrix{ \g_0 & 0 \cr 0 & \g_0}
\eeq
which enters in the definition of $\bar{\Psi}$. Consequently the
Lagrangian \rf{*6} is invariant under the generalized chiral
transformation \cite{aj}
\beq{*9}
\Psi \to e^{i \th \tGf/2 } \Psi,~~~~~\bar{\Psi} = e^{i \th\tGf/2} \bar{\Psi}.
\eeq
The symmetry expresses  nothing but the fermion number
conservation of the electroweak theory. In fact the current associated
with the symmetry is:
\beq{*10}
J_\m^5 = \bar{\Psi} \g_\m \tGf \Psi = \bar{\psi}_D\g_\m\g_5\psi_D
-\bar{\chi}_D\g_\m\g_5\chi_D,
\eeq
or, expressed in terms of the original fields of the model:
\beq{*10a}
J_\m^5 = \bar{\psi}_L \g_\m \psi_L +\bar{\psi'}_L \g_\m\psi'_L+
\bar{\chi}_R \g_\m \chi_R + \bar{\chi'}_R \g_\m \chi'_R
\eeq
i.e. the fermionic current. As remarked in the introduction this current
is known to be anomalous:
\beq{*11}
\partial_\m J_\m^5 = \frac{g^2_w}{32\pi^2} F^a_{\m\n}\tilde{F}^{a \m\n}
\eeq
and leads to the famous non-conservation of baryon number in the
electroweak theory. The reason we introduce the generalized   chirality
dictated by $\tGf$ is that the analogy to ordinary abelian symmetry
and the level shifting picture will be more clear, as emphasized recently
by several authors \cite{aj,amn}, who proved that under certain assumptions
a transition from one vacuum to another, related by a large gauge
transformation, will result in a level shift.
It is our goal to try to verify this level shifting for
realistic high temperature configurations which do not necessarily
satisfy the assumptions used in the analytical proofs of the
level shifting and, in addition, to provide further evidence that
we really observe continuum-like configurations during the simulations.

\section{ Lattice Formulation}\label{sec-lat}
First consider the time-independent continuum Dirac equation in a
background temporal gauge-fixed field $A_i={ A}_i^a{\tau^a}$:
\beq{*3.1}
(i\alpha_i\partial_i+A_i\alpha_i+\beta m)\psi(x)=E\psi(x).
\eeq
Here the index $i$ runs over the three spatial directions,
the wavefunction $\psi$ is a gauge doublet (weak isospinor)  with four
spinor components acted on by the $4\times4$ hermitian Dirac matrices
$\alpha_i$ and $\beta$. These satisfy
\beq{*3.2}
\{\alpha_i,\alpha_j\}=2\delta_{ij}\;;\;\{\alpha_i,\beta\}=0\;;\;
\beta^2=1.
\eeq
For the purposes of this section it is again useful to
consider the chiral representation of the $\g$-matrices, which means that
the $\a_i$ and $\b$ matrices are represented by:
\beq{*3.3}
\alpha_i=\pmatrix{\sigma_i&\cr&-\sigma_i}\;;\;
\beta= \pmatrix{&1_{2\times2}\cr1_{2\times2}& },
\eeq
where $\sigma_i$ are the Pauli matrices.

On the lattice we will be using the staggered formulation
for fermion fields.

\subsection{The massless case}
Let us in this section consider the massless limit $m=0$. 
The lattice transcription of \rf{*3.1} reads
\beq{*3.4}
{i\over2}\sum_{i=1}^3\eta_i(x)[U_i(x)\chi(x+\hat{i})-
           U_i^\dagger(x-\hat{i})\chi(x-\hat{i})]=E\chi(x),
\eeq
where $\chi(x)$ is a {\it single\/} spin component isospinor field,
the $U_i$ are SU(2) valued link variables defined on the lattice links,
which may be parameterized as $\exp(iaA_i(x))$, where $a$ is the lattice
spacing, and the $\eta_i(x)$ are the Kawamoto-Smit phases defined as
$\eta_i(x)=(-1)^{x_1+\cdots+x_{i-1}}$. It is well known that in the
limit where the gauge field varies smoothly on the scale of the lattice
spacing, the continuum limit of \rf{*3.4} recovers two copies of \rf{*3.1},
which we interpret as describing two independent fermion species \cite{1,2}.
These species become coupled by terms of higher order in $a$ \cite{3}
(i.e. by momentum  modes from the outer half of the Brillouin zone \cite{4}).

For an arbitrary background link field there are important symmetries in
the spectrum. If we denote by $(\chi,E)$ the existence of an eigenmode
$\chi(x)$ with eigenvalue $E$, then
\beq{*3.5a}
(\chi,E)\Leftrightarrow((-1)^{x_1+x_2+x_3}\chi,-E),
\eeq
and
\beq{*3.5b}
(\chi,E)\Leftrightarrow(\tau_2\chi^*,-E),
\eeq
where the Pauli matrix $\tau_2$ acts in weak isospace. Clearly applying
both symmetries \rf{*3.5a} and \rf{*3.5b} implies
that each mode is doubly degenerate.
Compare with the symmetries of the continuum equation \rf{*3.1}:
\beq{*3.6a}
(\psi,E)\Leftrightarrow(\gamma_5\psi,E),
\eeq
where $\gamma_5\equiv i\alpha_1\alpha_2\alpha_3$, and
\beq{*3.6b}
(\psi,E)\Leftrightarrow(C\tau_2\psi^*,E),
\eeq
where $C\equiv i\alpha_1\alpha_3$ in the particular Dirac matrix
representation \rf{*3.3}. There is a third symmetry, which has no analogue for
the lattice equation \rf{*3.4}:
\beq{*3.6c}
(\psi,E)\Leftrightarrow(\beta\alpha_1\alpha_2\alpha_3\psi,-E).
\eeq
Symmetry \rf{*3.6a} holds only for $m=0$, whereas \rf{*3.6b} and \rf{*3.6c}
 hold for any $m$.

The matrix $\gamma_5$ defines the chirality operator in the continuum
formulation. In order to represent chirality in the lattice system, it
is necessary to project onto the flavor singlet sector of the
multiple-species algebra implicit in \rf{*3.4}; ie. we need to define an
operator ${\Gamma}_5=\gamma_5\otimes1_2$, where the second matrix operates
on a ``flavor space'' \cite{3,4}. For the four-dimensional Euclidean Dirac
operator, this projection has been implemented by Smit and Vink in their
programme to compute topological charge \cite{5}, and by Hands and Teper in
studies of chiral symmetry breaking \cite{6}. The basic algebraic steps are
given in \cite{3,4}. For our three dimensional problem the algebra is very
similar, and here we give simply the result:
\beq{*3.7}
\la\chi_n|\Gamma_5|\chi_n\ra = {i\over2^3}\sum_x\sum_{ijk=\pm1}(-1)^{x_2}
\chi_n^\dagger(x)\,{\cal U}_{111}\chi_n(x+i\hat1+j\hat2+k\hat3).
\eeq
Here $\chi_n(x)$ is the $n$th eigenmode, and ${\cal U}_{111}$ denotes the
average of the SU(2) link products over the six equivalent paths joining
the two sites, thus ensuring the expectation value is gauge invariant.
Notice there is an important difference with the continuum operator.
Since ${\Gamma}_5$ is a three-link operator,
${\Gamma}_5^2\not=1$; only in
the deep continuum limit will ${\Gamma}_5$ and $\gamma_5$ coincide. At
non-zero lattice spacing ${\Gamma}_5$ requires in principle a finite
multiplicative renormalization,
which may be estimated perturbatively \cite{7}
or numerically \cite{5}.

Now consider what happens at a level crossing, when an eigenmode crosses
zero. There are never any exact zero modes in a lattice simulation, so we
will refer to modes which approach zero as ``quasi-zero modes''. Using
the symmetries \rf{*3.5a}, \rf{*3.5b} and the definition \rf{*3.7},
 we can see that if there is
such a mode with energy $\varepsilon$ and chirality $\gamma$, then there
will also be a mode with energy $-\varepsilon$ and chirality $-\gamma$.
As we shall see in subsequent sections, a lattice level crossing has the
following signal -- a mode initially has energy $\varepsilon$ and
chirality $\gamma_a$; as the simulation proceeds $\varepsilon$ becomes
smaller, eventually reaching a minimum before increasing again. The
chirality $\gamma$ should {\it reverse\/} sign at this point,
eventually reaching a value $-\gamma_b$. Since each
lattice mode is doubly degenerate, we interpret this process as {\it
two\/} positive energy modes of chirality $+\gamma_a$ crossing zero to
become negative energy modes of chirality $+\gamma_b$, and, using
symmetries \rf{*3.5a}, \rf{*3.5b}, {\it two\/} negative energy modes of 
chirality
$-\gamma_a$
crossing zero to become positive energy modes with chirality $-\gamma_b$.
The total change in chiral charge is thus $\Delta Q=-2(\gamma_a+\gamma_b)$
In the continuum limit as ${\Gamma}_5$ approximates $\gamma_5$ we should
find $|\gamma|$ approaching 1. Thus we find $|\Delta Q|\simeq 4$, which
should be compared with the result for level crossing for the continuum Dirac
Hamiltonian of $|\Delta Q|=2$: the extra factor of two is a consequence
of species doubling for lattice fermions. We shall see in the following
sections how well these ideal criteria are met in actual lattice
simulations.

\subsection{Masses and Yukawa coupling}
The situation becomes more complicated once a fermion mass is introduced.
As argued in \cite{8}, the algebra \rf{*3.2} obeyed by $\alpha_i$ and $\beta$ is
identical to that of the hermitian Euclidean matrices $\gamma_\mu$ in a
covariant formulation, which requires a four-dimensional lattice in order
to be represented faithfully by staggered fermions. Therefore to
incorporate mass we introduce a second ``timeslice'' of link variables,
identical to the first, and couple it to the original one as if it were
displaced by one unit in a fictitious Euclidean time direction. The
lattice equation becomes
\beq{*3.8}
{i\over2}\sum_i\eta_i(x)[U_i(x)\chi(x+\hat{i})-
           U_i^\dagger(x-\hat{i})\chi(x-\hat{i})]
+ m\eta_4(x)\chi(x+\hat4)=E\chi(x),
\eeq
ie. as if it were formulated on a $N^3\times2$ lattice with periodic
boundary conditions assumed in the 4th direction. The phase factor
$\eta_4(x)=(-1)^{x_1+x_2+x_3}$. Whether we regard this extra degree of
freedom as an extra timeslice connected to the first by unit gauge
connections, or simply as an extra component of the wavefunction $\chi$
is a matter of taste: the result is that we have a system which in the
continuum limit now describes four species of four-component fermions.
There are the following symmetries in the spectrum:
\bea
(\chi,E)&\Leftrightarrow&((-1)^{x_1+x_2+x_3+x_4}\chi,-E);\label{9a} \\
(\chi,E)&\Leftrightarrow&(\tau_2(-1)^{x_4}\chi^*,-E).\label{9b}
\eea
As before, these symmetries ensure that each lattice mode is two-fold
degenerate. However, the fact that the modes with energy $-E$ have minus
the chirality of the mode with energy $E$ remains unaltered. It is also
important to note that the chirality operator \rf{*3.7} links lattice sites in
the {\it same\/} timeslice:
therefore chirality for eigenmodes of the
massive equation \rf{*3.8} is simply the sum of chiralities
for each of the timeslices considered individually:
\beq{*mass}
\la \chi_n | \Gf | \chi_n\ra = \la \chi_{1n} | \Gf | \chi_{1n}\ra +
\la \chi_{2n} | \Gf | \chi_{2n}\ra.
\eeq
In this equation $\chi_{1n} (x)$ is an eigenvector of \rf{*3.8}
projected onto the first timeslice, and
$\chi_{2n}(x)$ the same eigenvector projected onto the second
timeslice.
We shall see that once mass is introduced, the
minimum value of $|E|$ is $m$, and that the chirality as measured by \rf{*3.7}
tends to have opposite signs on each of the two timeslices, thus reducing
the overall expectation. This is in accordance with our continuum intuition
that a mass term couples fields of opposite chirality.

Finally we discuss the incorporation of Yukawa interactions with a
Higgs field, so that the standard model can be discussed. Once again we
begin with the continuum formulation: if we use the
explicit representation of the Dirac matrices \rf{*3.3} we may according
to the Lagrangian \rf{*6}  write the equations of motion (suppressing
the primed fields, which are not
essential to the formulation) as
\beq{*3.11}
\pmatrix{i\sigma_i D_i& M\cr  M^\dagger&-i\sigma_i\partial_i\cr}
\pmatrix{\psi_L\cr\chi_R\cr}=E\pmatrix{\psi_L\cr\chi_R\cr},
\eeq
where each entry is a $2\times2$ matrix in spinor space and a $2\times 2$
matrix in isospace. The resulting system is very similar to the Dirac
equation for {\it four\/}-component Dirac spinors (recall the comments following
\rf{*6}).
Now, to model a mass term in the original system, i.e. to
incorporate a prescription for the matrix $\beta$, recall that we doubled
the number of spinor components by introducing a second timeslice with
identical gauge interactions, and had the mass terms couple the two
slices. Here we will do the same, except that on the second timeslice,
corresponding to the $\chi$ components, there will be no gauge
interactions. Gauge invariance will be enforced as in the continuum by
coupling the Higgs fields to the $\psi$ fields on the original timeslice.
The system of lattice equations then reads
\bea
&{i\over2}\sum_i\eta_i(x)[U_i(x)\psi(x+\hat{i})-
           U_i^\dagger(x-\hat{i})\psi(x-\hat{i})]
+ \eta_4(x)M(x)\chi(x)=E\psi(x), & \label{*3.12a} \\
&{i\over2}\sum_i\eta_i(x)[\chi(x+\hat{i})-\chi(x-\hat{i})]
+\eta_4(x)M^\dagger(x)\psi(x)=E\chi(x),&  \label{*3.12b}
\eea
where $\psi$ and $\chi$ are now single spin-component gauge-doublet
fields. The following symmetries exist:
\beq{*3.13a}
(\psi,\chi,E)\Leftrightarrow(\eta_4\psi,-\eta_4\chi,-E),
\eeq
and, if $h_u=h_d$,
\beq{*3.13b}
(\psi,\chi,E)\Leftrightarrow(\tau_2\psi^*,-\tau_2\chi^*,-E).
\eeq
Once again, these symmetries ensure that each lattice mode is doubly
degenerate if $h_u=h_d$; if not, there will be an additional fine
structure in the spectrum.

Because of the symmetry of the spectrum about $E=0$, the system described
by the lattice Hamiltonian \rf{*3.12a}-\rf{*3.12b} is not truly chiral:
if it were there
would be spectral asymmetry, and true spectral flow would result when a
mode crossed zero. Of course, the reason the model is not chiral is due
to the species replication inherent to the lattice approach (we leave
aside the issue of the requirement for complex representations, since in
any case it is perfectly possible to introduce a ${\rm U(1)}_Y$ gauge
interaction in the lattice system). For every positive energy level
which crosses zero to become negative, there will be a corresponding
negative energy level crossing zero to become positive. The symmetries
\rf{*3.13a},\rf{*3.13b}  ensure that at each zero crossing
{\it four\/} SU(2) doublets are
involved: two cross one way and two the other. Thus there is no true
baryogenesis on the lattice, since for each unit of baryon 
number created there is
another one destroyed. Of course, in the continuum limit we expect the
species to become independent (in the quenched approximation), and thus
we should be able to focus simply on the crossing itself as the event of
interest. Note also that chirality is still defined by \rf{*3.7}, and that we
have the freedom to either define the total chirality of the mode
as $\la\psi|\Gamma_5|\psi\ra+\la\chi|\Gamma_5|\chi\ra$, or the more
sophisticated version $\la\psi|\Gamma_5|\psi\ra-\la\chi|\Gamma_5|\chi\ra$.
As we saw in sect. \ref{sec-mod}
it is the last version which is directly related to
the anomalous fermion number of the electroweak theory and we will
use this definition in the following:
\beq{*G5}
\la \Psi | \tGf |\Psi\ra \equiv \la \psi | \Gf |\psi\ra-
\la \chi | \Gf | \chi\ra, ~~~~~\Psi \equiv \pmatrix{\psi \cr \chi}
\eeq

\section{Calibration}\label{sec-cal}

As mentioned above there seem not to be any analytical results
which tell us in detail about the level crossing for chiral fermions
coupled to the gauge-Higgs system encountered in the standard
model. In order to know what to expect we have investigated this
question numerically on the lattice by taking families of smooth
configurations\footnote{By ``smooth'' configurations we  mean
lattice configurations which are approximations to continuum configurations.}
which interpolate between different classical
vacua and for these configurations we have followed the flow
of eigenvalues and the change of chiralities. We have investigated
the following scenarios:
\begin{itemize}
\item[(1)] A family of gauge-Higgs  field configurations
which interpolate between two vacua and where the zero of the Higgs field
appears at the same time as the Chern-Simons number
$N_{cs} (t)$ of the gauge field is equal to 1/2.
\item[(2)] A family of field configurations  where the gauge- and
Higgs fields still interpolate between two vacua, but where the zero of the
Higgs field is displaced in time relative to the time $t$ where $N_{cs}=1/2$.
\item[(3)] A family of field configurations where only the gauge field
interpolates between two neighboring gauge vacua,
while the Higgs field remains constant.
\item[(4)] A family of field configurations where only the Higgs field
interpolates between  two neighboring Higgs vacua,
while the gauge field stays constant (for instance zero).
\end{itemize}
The field configurations are constructed as follows: Let us first define
a gauge transformation $V(x)$ with winding number one:
\beq{cal1}
V(x) = (-1)\exp\left(\frac{ 2\pi  i \sg_i v_i}{L |v(x)|}
\max \{ |v_j (x) |, j=1,2,3\} \right),
\eeq
where $x$ denotes a lattice point, $L$ the linear size of the lattice,
$v_i = x_i -\oh L$ and $\sg_i$ are the Pauli matrices.
The trivial vacuum is given by
\beq{cal2}
U_i^{(0)} (x)=\hat{1},~~~~~~\phi^{(0)} (x) = \pmatrix{0 \cr 1},
\eeq
while a vacuum configuration with winding number one is given by
\bea
U_i^{(1)}(x) &=& V(x)V^{-1}(x+\hat{e}_i)=\exp (i\sg_i u_i(x))
\label{cal4} \\
\phi^{(1)} (x)& =& V(x) \phi^{(0)}.  \label{cal3}
\eea
where $\hat{e}_i$ is a unit lattice vector in direction $i$ and
$u_i(x)$ will be close to zero for large lattices.

A continuous family of configurations which interpolate between the
two vacua is simply given by
\beq{cal5}
U_i (x,t) = \exp (if(t) \sg_i u_i (x)),
\eeq
\beq{cal6}
\phi(x,t) = (1-g(t))\phi^{(0)}+ f(t)\phi^{(1)} (x),
\eeq
where $f(t)$ and $g(t)$ are monotonic, continuous functions which
interpolate between 0 and 1 when $t$ changes from 0 to the final
time which we  denote $T$.

Note that there exists precisely one
space-time point $(x,t)$ where $\phi(x,t) =0$ and the time $t$ is
determined by $1-g(t) = f(t)$. We denote this time $t_h$.
If $g(t)=f(t)$ it will coincide
with the time $t_{1/2}$ where $N_{cs}(t) =1/2$, but if $g \neq f$
it will in general be different from $t_{1/2}$.

We will choose $f(t) = t/T$ in the actual numerical calculations.
This means that $t_{1/2} = T/2$. In case (1) and (4) mentioned above we choose
$g(t)=f(t)$.  In case (2) we choose $g(t)= \min \{ 3t/T,1\}$, which means
that the zero of the Higgs field will be located at $t =T/4$, ($t_h=T/4$)
i.e. displaced by $T/4$ relative to $t_{1/2}$.

Let us now describe the spectral flow and the change in chirality for
the four families of field configurations mentioned above.
In fig. \ref{figcal0} we show the Chern-Simons number as
a function of the time $t$ for the gauge fields given by \rf{cal5}.
In addition we show in fig. \ref{figcal1} 
the change in the lowest eigenvalue and the change
in chirality for various values of the Yukawa coupling. Everything
behaves as expected from the continuum considerations.

In case (2) where the zero of the Higgs field is displaced from $t_{1/2}$
we see that the time $t_0$ at which the lowest eigenvalue crosses zero
changes continuously with increasing Yukawa coupling from $t_{1/2}$,
corresponding to the crossing at $N_{cs} =1/2$ for zero Yukawa coupling,
towards $t_{h}$, which is the time where the Higgs field has a zero.
This is illustrated in fig. \ref{figcal2}. The flip in chirality
always occurs at $t_0$, as expected.

In case (3) where the Higgs field is constant we see that the
level crossing is still present for moderate values of the Yukawa
coupling. This is in sharp contrast to the situation where
we insert an explicit mass term in the Dirac equation. In that case we
know from the continuum equation that there is a mass gap, separating the
negative and the positive eigenvalues, and this is also seen explicitly
in the lattice implementation (see sec. \ref{sec-mea}).
When the Yukawa coupling
is increased $t_0$ is gradually displaced towards $T$, and for a finite
value of the Yukawa coupling $t_0$ reaches $T$ and there is no longer
any level crossing and no chirality flip.
This scenario is shown in fig. \ref{figcal3}.

The behavior of the lowest eigenvalue in the situation where the
gauge field is zero (i.e. $U_i(x)$ is given by \rf{cal2}) and
Higgs field is given by \rf{cal6} with $g(t)=f(t)= t/T$ is shown
in fig. \ref{figcal4} as a function of the Yukawa coupling. The situation
is in a certain sense dual to the one in case (3). For small Yukawa
coupling there is no level crossing, but eventually a zero and
a corresponding chirality flip takes place at $T$ and with increasing
Yukawa coupling $t_0$ moves towards $t_h$.
\vspace{24pt}

The results in this section are relevant when we want to interpret
the data obtained in real lattice simulations. In particular we
see that increasing the Yukawa coupling can displace the time $t_0$ of
zero level crossing. It is somewhat surprising that there is no
need for a zero in the Higgs field in order to observe level crossing,
provided the Yukawa coupling is sufficiently small. However,
our three dimensional lattice has periodic boundary conditions and since
it is finite, it imitates a compact space of finite volume.
This means that  the ``continuum'' configuration imitated by
\rf{cal5} for the gauge field and \rf{cal2} for the Higgs field
will have finite energy, and
we expect  spectral flow of the continuum Dirac operator
to be continuous with respect to the
coupling constants in the lagrangian. Since we have level crossing
for zero Yukawa coupling by the Atiah-Singer index theorem
and since the chiral symmetry is not broken for finite Yukawa coupling,
we must have level crossing for small Yukawa couplings as well, even if the
Higgs field is nowhere zero. This is indeed what we observe. For sufficiently
large Yukawa coupling the level crossing can disappear without violating
any continuity requirements by the mechanism shown in fig. \ref{figcal3}.
It is more difficult to decide whether these continuity arguments remain
valid if the three dimensional space is non-compact. The very simple
configurations we have used will have infinite energy in the non-compact case
and might not belong to a reasonable class of field configurations
where one can define a spectral flow of the Dirac operator
$H(A_i(t), \phi(t))$. Similar remarks are valid in case (4).
On the other hand the results obtained in case (2) are probably
true also for a non-compact continuum space, even if if the actual
configuration used here has infinite energy in the continuum limit
for $t\neq 0,T$. The reason is that it is easy to find
finite energy continuum field configurations with the characteristics
used in case (2).
An analytic proof of the behavior observed in case (2) has recently
been given if space is one dimensional \cite{ajn}.
It would be interesting to extend it to three dimensions.

At high temperature we might certainly encounter situations
where the phase of the Higgs and the gauge field are placed in different
gauge sectors as in case (3) or (4),  since the finite temperature
provides the needed infinite energy. Consequently cases (3) and (4) are
relevant for our interpretation of data not only because the
spatial lattice used in the simulations is finite, 
but also because the situations might
be generic at the high temperatures prevailing in the early
universe above the electroweak phase transition.

\section{The numerical method}\label{sec-num}

\subsection{The approximation}
The following approximations have been used in this work:
we consider only one weak isospinor fermion and the two associated singlet
fermions and we ignore the hypercharge sector. These approximations
 are presumably not important for the questions we want to
address. In addition we completely ignore the feedback of
fermions on the gauge and Higgs fields and we always treat the gauge
and Higgs fields as background fields when we ask dynamical questions
about the fermions. This approximation is difficult to control, but
it might still not be very important for the dynamics of the gauge and
Higgs fields. The basic approximation, when it comes to the dynamics
of the gauge and Higgs fields, is the  use of classical
physics as the tool for simulating  the high temperature fluctuations in
the electroweak theory.
One would expect this approximation to give a  reliable representation
of the magnetic sector of the theory since the classical partition
function on the lattice\footnote{Let us remind the reader that there
exists no classical partition function of a continuum field theory due
to the Rayleigh-Jeans instability. If one restricts the number of degrees
of freedom per unit volume by considering a lattice version of the
field theory the concept of a classical partition function is perfectly
well defined.} is identical to the naive infinite temperature limit
of the full quantum partition function. If we discuss the electric
properties of the theory it is by now well known that we need to
supplement the naive dimensional reduction with mass and charge
renormalizations and in addition with Higgs fields in the adjoint
representation (the remnants of the $A_0$ component)
in order to get a correct representation of the
full quantum theory \cite{reisz}. The Gauss constraint will
be satisfied by the use of the classical equations, and this
means that the kinetic part of the adjoint Higgs field coupled to
the gauge field {\it will} be included in the classical approximation
(The Gauss constraint is equivalent to a term $(D_i^\dg A_0)^a (D_i A_0)^a$ in
the Lagrangian). However, there will be additional terms like $A_0^2,
A_0^4$ and $A_0^6$ in the effective three-dimensional
high temperature expansion of the
action, where especially the first two terms are important for
the Debye screening mass. We will ignore these subtleties
here\footnote{The role of these additional finite temperature
corrections is not clear for the kind of questions we will deal with.
Leaving aside the question of a gauge invariant
meaning of the Debye screening mass,
it is a static quantity, calculated as a temperature average in a
canonical ensemble, while we consider dynamical processes such as
sphaleron transitions which occur in real time.}
and view the present computer simulations as generating
typical high temperature fluctuations, the temperature being
judged to correspond to fluctuations in the unbroken phase since
$\la \vf^2 \ra$ is significantly smaller than its tree value
for the choice of lattice coupling constants we will use (see later).

Classical thermal fluctuations can be generated either by using
a canonical ensemble and applying some standard Monte Carlo updating
like heat bath or the Metropolis algorithm to the field configuration,
or (more fundamentally) to use a micro-canonical ensemble and just
let it develop according to the classical equations of motion. In the
first case the temperature is kept fixed, while the energy per degree
of freedom will have fluctuations proportional to
$1/\sqrt{N}$, $N$ being the number of degrees of
freedom. In the second case the energy will be constant, while the
temperature of the system will fluctuate as $1/\sqrt{N}$. From
a numerical point of view the first method is preferable, since
the second is known to suffer from problems with ergodicity for
too small systems. We have nevertheless chosen to use the classical
equations of motion since they reflect very directly the real time
evolution of the field configurations. However, in order to use the classical
equations of motion we have to prepare an initial distribution of fields
and their conjugate momenta corresponding to a given temperature, and
here we have to use the heat bath or the Metropolis algorithm. For
systems with gauge invariance this means additional complications
since the Gauss constraint has to be
satisfied for the classical configurations
and this will not be the case for a Metropolis updating. This technical
obstacle can be circumvented by introducing the Gauss constraint as a
Lagrange multiplier (we refer to \cite{aaps} for details).
After the initial thermalization by means of Metropolis,
 we let the system develop according
to the classical equations, discretized in a way which
automatically conserves the Gauss constraint (see next subsection).
For sufficiently large systems
the fundamental hypothesis of ergodicity of the micro-canonical
ensemble will ensure that the phase space surface of
constant energy is covered uniformly if the system evolves according
to the classical equations of motion.

\subsection{The gauge-Higgs system}

As mentioned above the sequence of
gauge and Higgs field configurations we are using
is generated by the classical equations of motion.
By that we mean the following: we choose the standard  Wilson action
for Euclidean lattice gauge theory combined with the standard Higgs action
on a lattice and make the obvious change of sign of the spatial part
relative to the temporal part in order to get a Minkowskian signature.
After that we make the lattice spacing in the temporal direction much smaller
than in the spatial directions by scaling the lattice spacing ``$a$'' to
``$a \Delta t$''. In this way the action will be split
into a kinetic and a potential part:
\beq{*4.1}
S = \sum_t (E_{kin} - E_{pot})
\eeq
where the summation is over all time slices. From $S$ one can derive the
classical equations of motion which connect timeslice $t$ and $t+\Delta t$.
These equations of motion will automatically satisfy the Gauss constraint.

If we define an energy functional $H$ by
\beq{*4.2}
H= E_{kin}+E_{pot}
\eeq
it will not be strictly conserved by the equations of motion since
time is still discrete. However, in the limit $\Delta t \to 0$, $H$ reduces
to the correct Hamiltonian of the system and the energy will be conserved
by the equations of motion. For $\Delta t \leq 0.05$ we have, for the range
of coupling constants used in the simulations, found very good conservation
of the energy.

For a detailed
discussion of the equation of motion derived from
\rf{*4.1} we refer to \cite{aaps}; let us only here give explicitly
the form of $\sum_t E_{kin}$ and $\sum_t E_{pot}$ used in the simulations:
\bea
\sum_t E_{kin} &= & \frac{\b_G}{(\Delta t)^2}
    \sum_{\Box (t)} (1-\oh \Tr U_{\Box (t)}) +\nonumber \\
&&   \frac{\b_H}{2(\Delta t)^2} \sum_{x,t} \Tr \left(\Phi^\dg_x \Phi_x -
\Phi_x^\dg U_{x,x+\hat{0}}\Phi_{x+\hat{0}} \right)    \label{*4.3a}\\
\sum_t E_{pot} &= &  \b_G \sum_{\Box (s)} (1-\oh \Tr U_{\Box (s)})
+ \frac{\b_H}{2} \sum_{x,t,i} \Tr\left( \Phi^\dg_x \Phi_x -
   \Phi_x^\dg U_{x,x+\hat{i}} \Phi_{x+\hat{i}}\right)\nonumber \\
&&
  +\b_R \sum_{x,t} (\oh \Tr \Phi^\dg_{x,t} \Phi_{x,t} -v^2 )^2
\label{*4.3b}
\eea
where $U_{x,x+\m}$ is the lattice gauge field on the link
connecting $x$ and $x+\hat{\m}$,
 $U_\Box$ denotes the plaquette action for the gauge field and
$\Box (t)$ stands for a plaquette in the $\hat{0}-\hat{i}$ plane while
$\Box (s)$ refers to a plaquette in a $\hat{i}-\hat{j}$ plane. 
The $SU(2)$ Higgs doublet $\vf$ is represented as a matrix
\beq{*4.4}
\Phi \equiv
\left( \begin{array}{cc} \phi_0^* & \phi_+ \\
                         -\phi_+^* & \phi_0
       \end{array}
\right).
\eeq    
The
summation over spatial points is finite and limited by the lattice
3-volume (periodic boundary conditions)
while it is infinite in the time direction.

The Metropolis updating used to thermalize the system uses as weight
\beq{*4.2a}
\exp (-H_\xi),~~~~H_\xi = H + \xi G^2
\eeq
where $H$ is the lattice hamiltonian given by \rf{*4.2}, while $G$ denotes
the lattice version of Gauss constraint (see \cite{aaps} for details).
By choosing $\xi$ sufficiently large one can (at the expense of slowing down
the thermalization time) control the violation of Gauss constraint to be
arbitrarily small in the Metropolis updating.

If we write the classical continuum Hamiltonian as
\beq{*4.5}
H = \int d^3 x \left[ \oh E^a_iE^a_i + \oq F_{ij}^aF_{ij}^a +
|\pi|^2+|D_i \vf|^2 + \l (|\vf|^2-v^2_c)^2\right]
\eeq
 the tree value
connection between the lattice parameters in \rf{*4.3a}, \rf{*4.3b} and the
continuum coupling constants in \rf{*4.5} is as follows\footnote{We
keep this unnecessary complicated notation for historical reasons}
\beq{*4.6}
M_c^2  =  \frac{2(1-2\b_R -3\b_H)}{\b_H a^2},~~~~~
\l_c  =  \frac{8\b_R}{\b_H^2 },~~~~~
g^2_c  =  \frac{4}{\b_G }
\eeq
where $a$ denotes the lattice spacing and
\beq{*4.7}
v^2_c a^2 =\frac{\b_H}{2}  v^2 =\frac{\b_H}{4\b_R}( 2\b_R+3\b_H -1).
\eeq

The naive continuum limit $a \to 0$
imposes a fine tuning on $(2\b_R + 3\b_H)$.
If we ignore thermal fluctuations and determine the $W$-mass $M_W$ and
the Higgs mass $M_H$ from the coupling constants  in the Hamiltonian we have:
\beq{*4.7a}
M_H^2a^2 = \frac{16\b_R}{\b_H} v^2,~~~~~~~M_W^2a^2 = \frac{\b_H}{\b_G} v^2.
\eeq
If for convenience, to restrict the coupling constant space, we impose
 the constraint $M_W = M_H$, we get the relation
\beq{*4.7b}
\b_R = \frac{\b_H^2}{16 \b_G}.
\eeq
From \cite{aaps} we know that a reasonable choice is  $\b_G \geq 10$ and
$\b_H$ in the neighborhood of $1/3$. This means that $\b_R$ is quite small.

\subsection{The procedure}
Equipped with an initial configuration at a given temperature and
the classical equations of motion we can now follow the
evolution of the \CSn by measuring the integral
\beq{*4.8}
Q(t) \equiv N_{CS}(t)-N_{CS}(0) =
\frac{1}{32 \pi^2} \int_0^t dt \int d^3 x\; F^a_{\m\n}F^{a\m\n}.
\eeq
This quantity has a simple implementation on the lattice and will
directly give us the change in 
Chern-Simons number, which is a gauge invariant
quantity. This procedure has been applied succesfully before
(see \cite{aaps} for details).  The typical measurement of $Q(t)$ will
result in a curve which stays for some time around integer values
of $Q(t)$, while the transitions between these values take place
rather rapidly. Superimposed on these movements of $Q(t)$ will be
short wavelength thermal fluctuations,
which occasionally can mask somewhat the picture
of $Q(t)$ jumping between integer values. If desired, it is possible to
strip off these thermal fluctuations in the following way. By solving
the classical equations of motions we get a time sequence of configurations.
For each of these configurations we can apply the simplest relaxation
equation:
\beq{*4.9}
\frac{\prt \vf}{\prt t} = -\frac{\dl H}{\dl\vf},~~~~~~~
\frac{\prt A}{\prt t} = -\frac{\dl H}{\dl A}
\eeq
where $H$ is the Hamiltonian \rf{*4.2}. This technique is well known from the
lattice study of monopoles and instantons  \cite{6,8,bt,ilmss,ls}.
If one ``cools'' the configurations  too much one forces them
to a vacuum configuration, which is not what we want. However,
only a few cooling
steps will reduce the energy of the gauge
configurations by a factor 80 without
breaking the link to the original configuration since the part of the field
configurations which survives this relaxation is precisely the long
wavelength fluctuations which we imagine are responsible for the
large scale changes of $Q(t)$. Indeed, the picture of the transitions
between different integer values of $Q(t)$ is considerably sharpened
by such a cooling. This effect is illustrated in fig. \ref{fig1}.

In order to check for level crossing during the time-development of the
gauge field configurations we calculate the eigenvalues of the Dirac
Hamiltonian for each time step, using the staggered fermion formalism
discussed in detail in sections \ref{sec-mod} and \ref{sec-lat}. 
Since  the Dirac operator
$H_D(A(t), \vf(t))$ will depend implicitly on the time $t$ due to the
time dependence of the gauge and Higgs field we  will get a spectral
flow of the eigenvalues and we can compare this with the expectation
that a crossing of zero should be related to a change in \CSn and
that diving and rising eigenvalues have different generalized chiralities.

 The spectrum of the Dirac operator was found numerically by first using the
Lanczos algorithm to tridiagonalize the hermitian Hamiltonian matrix, and then
using Sturm bisection to extract the eigenvalues. As described in \cite{6},
on large systems this procedure finds eigenvalues iteratively, with
convergence tending to occur first for extremal values of $|E|$. We
found that for the systems we investigated the smallest eigenvalue
converged after roughly 300 iterations of the Lanczos algorithm, and the
lowest ten eigenvalues after roughly 700 iterations.
Once the eigenvalues had converged the eigenmodes of the tridiagonal
matrix are easily extracted by inverse iteration: the Lanczos algorithm is then
rerun with the same initial vector in order to perform a unitary
transformation on the eigenvectors of the tridiagonal matrix to convert
them to eigenvectors of the original Hamiltonian. We
checked that the resulting residual vectors had norms in the range
$10^{-6}$ -- $10^{-10}$, which is adequate for the subsequent
measurements of chirality.

\section{The measurements}\label{sec-mea}
\subsection{Results for the massless Dirac operator}
In \cite{af} the spectral flow of the {\it massless}
Dirac operator\footnote{By ``massless'' we mean that the mass is explicitly
chosen zero and that there is no coupling to the Higgs field.}
was investigated along the lines outlined above. It was shown that
there indeed was a crossing of zero associated with a sphaleron-like
transition where the \CSn changed by one unit. Some questions were
left unanswered, however. 

The first question was connected with the
chirality. As explained in sect. \ref{sec-lat} the spectrum is symmetric with
respect to energy $E=0$. If an eigenvalue of chirality +1 crosses
zero from above, a corresponding eigenvalue of chirality -1 should
cross zero from below. This is indeed what we observe. In fig. \ref{fig1}
we have shown the change in \CSn which characterizes a typical
sphaleron-like transition. In fig. \ref{fig2} we show the corresponding
diving of the lowest eigenvalue and the associated change in chirality.

The second question was
related to the observation that a number
of configuration histories had a \CSn which changed by
1/2 and then returned to its original value. These changes survived
relaxation and they should consequently have some topological
content. This hypothesis was supported by observing that their
appearance was accompanied by a diving of eigenvalues to zero.
It was conjectured  in \cite{af} that one could view such configurations as
gauge-Higgs configurations which climbed to the sphaleron barrier
where the \CSn is 1/2, but then rolled back down to the original vacuum
rather than to the neighboring one. By measuring the chirality we can
substantiate these conjectures. This is illustrated in fig. \ref{fig3},
where we
show such a situation. The eigenvalues vanish twice, the two zeroes
being quite close, while the chirality
at the first zero changes from +1 to -1 and back again to +1 at the
second zero. The obvious interpretation is that the field configurations
reach a ``sphaleron''-like configuration (where the energy is zero)
and that it also overshoots it
a little causing the negative chirality state to cross zero,
but afterwards it returns to its original vacuum in agreement with a
total change of \CSn of zero, and once again the chirality states cross zero.

We conclude that the gauge fields seem to qualify as continuum
configurations. The change in \CSn is reflected closely in the
spectral flow of the eigenvalues of the massless Dirac operator,
which is implemented by means of staggered fermions.

\subsection{Massive fermions}
Before we turn to the ``complete'' electroweak theory, where
the fermions acquire their masses via the Higgs field, it is
illuminating to address the coupling of the gauge field to the
massive fermions as discussed in sec. \ref{sec-lat}.
In this case the continuum
fermion  is a four-component Dirac spinor and has isospin 1/2.
Recall that the lattice implementation uses two fields
with support on alternating time slices to represent
the lattice version of the Dirac matrix $\b$, and chirality
for a given eigenmode is defined by \rf{*mass}\footnote{It makes no sense
to use the generalized chirality $\tGf$ since  the spinors in this case
do not split into different isospin representations}.

In this case the eigenvalues of the Dirac operator do not cross zero
if solved in a sequence of gauge field configurations where the
\CSn changes by one. In fig. \ref{fig4}a we show the lowest eigenvalue for a
number of different masses and we see that the lowest eigenvalue
is approximately equal to  the mass $m$, in agreement with expectations
from the continuum formalism.
We have studied the chirality of the lowest mode as a function of the
mass. Some results are shown in fig. \ref{fig4}b.
If the mass is zero the component $\chi_{2n}(x)$ (see \rf{*mass}) on
the second timeslice does not  contribute to the chirality as defined
by $\rf{*mass}$. However even for small masses both components will start
to contribute, but the sum for small masses is quite similar to that
obtained in the massless case. However, as the mass increases
$\la \Gf \ra$ decreases towards zero, due to a cancellation between
the contributions from the two timeslices. For the lowest eigenmode
it seems still possible to observe a change in sign of $\la \Gf \ra$
even for large masses. If we look at higher eigenmodes the
chirality is closer to one and is less affected by the mass
term. All this is in agreement with expectations from the
continuum theory and
verifies that the ``two-timeslice'' formalism works satisfactorily.
We are now ready to apply the formalism to the electroweak theory.

\subsection{Spectral flow of the full Dirac operator}

In the following we will assume, unless explicitly mentioned, that
$h_u=h_d$ and we will denote the common coupling constant by $h_y$.
We have investigated the level crossing picture and the change in the
generalized chirality $\tGf$ for a number of sequences of configurations,
each sequence corresponding to a change in \CSn of order one. For each of
these sphaleron-like transitions the eigenvalues and  eigenmodes
of the Dirac equations \rf{*3.12a},\rf{*3.12b}  were found for
different values of the Yukawa coupling $h_y$ ranging from 0.05 to 1.
In addition these measurements were performed for three different
``versions'' of the same sphaleron-like transition, corresponding
to different degrees of relaxation of the individual configurations,
more precisely 0, 4 and 8 relaxation steps in the discretized version
of eq. \rf{*4.9}, which had the effect of stripping off as much as
96 \% of the short wavelength thermal energy in the case of 8
relaxation steps.

The picture is very consistent for the various transitions and we will
present  details from two  of them. In fig. \ref{fig5}a and \ref{fig6}a
we have shown the lowest eigenvalue for the two sequences of
configurations (parametrized by the time $t$)
near sphaleron transitions for Yukawa couplings
$h_y= 0.1$, 0.25 and 0.5 (fig. \ref{fig5}a) and $h_y=0.1$ and 0.5 
(fig. \ref{fig6}a).
In fig. \ref{fig5}b and \ref{fig6}b we
have shown the measured generalized chirality $\avGf$ for the same sequences
of configurations and the same values of the Yukawa coupling. 8 relaxation
steps have been imposed on the configurations.
The next to lowest eigenvalues are only weakly affected
and the same holds for the  corresponding
$\avGf $ values and we have chosen not to show these.

The effect of the relaxation is to
increase the value of $\avGf$ slightly, and to affect only rather weakly
the lowest eigenvalues in the case where $h_y=0.1$. If $h_y= 0.5$ the
effect of cooling is somewhat stronger. This is show in fig. \ref{fig7}
for the transition shown in fig. \ref{fig6}. In fig. \ref{fig7}a
and \ref{fig7}b $h_y=0.1$ and three curves are shown of the lowest
eigenvalue and $\avGf$, respectively, corresponding
to 0, 4 and 8 relaxation steps. The lowering of $\avGf$ in the
case of no relaxation is in accordance with the discussion in sec.
\ref{sec-lat} and is presumably caused by fluctuations of the Higgs field.
In fig. \ref{fig7}c and \ref{fig7}d we  show the same set of
curves, but generated for $h_y =0.5$.
It is clear from the figures that it will
be difficult to conclude anything about level crossing and
chirality flip of $h_y > 0.5$ without relaxation of the individual
configurations. When $h_y \geq 1$ it is impossible even with  extensive
relaxation. For such large values of $h_y$ there seems to have
been introduced a lot of spurious
transitions due to the fluctuations of the radial part of the Higgs field.

It is of interest to understand the relationship between the
Higgs field and the transition between positive and negative
chirality. As we increase the Yukawa coupling we indeed observed
some change in the precise location of the ``zero'' energy
solution\footnote{As remarked previously a ``zero'' energy mode
should be understood as a quasi-zero mode. The discrete approximations
used in the numerical calculations imply that we have to be content
with eigenvalues which get (very) close to zero.} as is clear from
fig. \ref{fig5}b and \ref{fig6}b. The change is
small compared to the total width of the sphaleron-like transition
(the region where the \CSn change by approximately one unit), but
it nevertheless shows that there cannot be a strict relation
between the zeros of the Higgs fields and the occurrence of
a zero mode. This is in complete agreement with the analysis performed
in section \ref{sec-cal}, where we used artificial ``smooth''
configurations. On the other hand we have to say that our setup is
not the best for an investigation of such a relationship. The reason
is that we are in the symmetric phase of the electroweak theory.
The expectation  value of the Higgs field is quite small and it fluctuates
significantly. This means that $\phi(x,t)$ might get close to zero in
a number of places and the relaxation we use is insufficient to change the
expectation value of the Higgs field very much. The phases of the Higgs
and the gauge field seem to align much faster. This is due to the
small value of the Higgs coupling which is dictated by the formal
continuum limit. It would be interesting to perform the same measurements
on configurations where the temperature is below
the transition point, such that we are still in the broken phase.
Work in this direction is in progress.

Finally we have investigated the situation with different Yukawa couplings
for the upper and lower components ($h_u = 0.1$ and $h_d = 0.25$).
In this case the Dirac Hamiltonian ceases to be charge conjugation symmetric,
and as discussed in sec. \ref{sec-lat} we expect states
which were previously degenerate to
split into two states of different energy.
To test this we calculated the inner product
$<\Psi_i\mid\eta_4\tau_2\Psi_j^*>$, where $i$ and $j$ denote the two eigenmodes
of nearly equal energy. According to the symmetries
\rf{*3.13a}, \rf{*3.13b}, the previously degenerate eigenmodes would have
been related as $\Psi_j\equiv\eta_4\tau_2\Psi_i^*$ in the limit $h_u=h_d$.
As expected, we find the inner product to be close to
unity for all such pairs.

\section{Discussion}\label{sec-dis}

The purpose of this work has been to establish the  possibility to investigate
in detail by numerical methods a number of questions related to
sphaleron-like transitions in the early universe.
By simulations we have generated conditions which should have some
resemblance with the conditions near the electroweak phase transition:
a small expectation value of $|\vf|^2$ (compared to the tree-value
determined from the Lagrange function), but the temperature still
not so high that it completely masks the transitions between neighboring
vacua. We are here in the lucky situation that we can use our finite
lattice volume in a constructive way. Had we been working at infinite
or very large volume $V$ it is clear that the temperature
fluctuations in globally defined quantities like the \CSn would be very large
since they essentially grow with the square root of the volume.
By fine tuning the volume and the coupling constants we can work at
high temperature and corresponding small expectation values of $|\vf|^2$,
but still have an acceptably small change in the \CSn. The risk is that
finite volume artifacts will be present, and certainly we would have
to check this aspect carefully if we attempted to measure actual
transition rates. Then one would have to make sure than the transition
rate grew correctly with volume. In this study we have concentrated
on the qualitative aspects of the individual transitions: is there any
trace of topology left in the high temperature phase above the
electroweak transition and can it be associated with level crossing in
the usual way?

We have seen that the gauge fields indeed behave like continuum configurations
with respect to  change  of \CSn and  level crossing and chirality of
the massless fermions. As mentioned in the introduction it is not
a priori clear how this picture survives the coupling to the
Higgs sector of the electroweak theory.
However, for not too large values of the Yukawa coupling
$h_y$ ( $h_y \leq 0.5$ ) we have observed the diving of a single
eigenvalue to zero and an associated flip in the generalized
chirality $\tGf$ when the \CSn changes by one unit.
The natural interpretation of this is that energy eigenvalues of
eigenmodes of opposite generalized chirality cross at zero energy.
The crossing is not linked in any precise way to the point where
$\Delta N_{CS} = 1/2$, but as discussed in sec. \ref{sec-cal} one
would not expect this for  general gauge-Higgs
configurations\footnote{Even for Yukawa coupling zero we are not
aware of any analytic considerations which tell us that the crossing
has to take place at the point where $\Delta N_{CS}=1/2$. Furthermore
our configurations do not interpolate between two vacuum configurations
due to the finite temperature. The best we can say is that we start out
in some neighborhood of a classical vacuum and end up near another
classical vacuum.}.
The observed picture of  level crossing and especially
chirality flip  is enhanced  by stripping off the high energy
thermal fluctuations of the gauge and the Higgs fields.
This is an indication that the low frequency part of the Higgs field
indeed has some topological content, even if it is clear that the
Higgs configurations from the outset are rather ``wild''. It also
shows that we presumably have a ``renormalization'' of the chirality
due to fluctuations, as mentioned in sec. \ref{sec-lat}. The break down
of any sensible  eigenvalue flow and  chirality flip for $h_y \geq 1$
is probably due to the fluctuations of the radial part of the Higgs
field. In sec. \ref{sec-cal} where the radial part of the Higgs
field was smooth we could choose a considerably larger $h_y$ before
we encounted any problems with eigenvalues and chirality. When we
use the relaxation algorithm on the configuration the situation
does not improve much. On the other hand we know from direct
measurements that the phase of the Higgs field and the gauge field
are almost totally correlated after 8 relaxation steps and that
we get very sensible results for smaller values of $h_y$. Since
our relaxation algorithm works rather slowly on the radial part
of the Higgs field, it is natural to expect the fluctuations of
the radial part to develop into lattice artifacts for some
threshold value of $h_y$. This threshold value should depend somewhat
on the relaxation and we have observed this dependence.

Finally we wanted to use the measured flow of eigenvalues
and the chirality flip to extract information about the topology
of the Higgs field. The results obtained
in sec. \ref{sec-cal} made this procedure more
ambigious than desired. Before relaxation we can by direct
measurement see that the phase of the Higgs field and the
gauge field are rather decorrelated. After relaxation they are
highly correlated. The basic picture of levelcrossing and chirality
flip is not changed by relaxation for $h_y \leq 0.5$ although it
becomes much clearer. Unfortunately we have seen in sec. \ref{sec-cal}
that we can have level crossing and chirality flip even if the
phase of the gauge and the Higgs field  develop into different
vacua. This means the the cooling itself could induce a change
of the relative phase of the gauge and the Higgs field
(from being in different vacuum sectors to being in the same vacuum sector)
without a drastic change in the levelcrossing and chirality flip.
To determine in more detail the topology associated with  the Higgs field
would therefore require additional independent measurements.

\vspace{12pt}

{\bf Acknowledgements} - SJH was partly supported by a CERN Fellowship
and is now supported by a SERC Advanced Fellowship.
KF would like to thank the CERN Theory Division for hospitality during
the final phase of this work. GK thanks DESY and  NBI for their
warm hospitality. JA and GT thank the Danish Research Coucil for
a grant which made it possible to run part of the simulations on
the UNI-C CM-machine.

\vskip 0.5 truecm

\addtolength{\baselineskip}{-0.30\baselineskip}

\newpage
\begin{center} {\bf Figure Captions} \end{center}

\begin{figure}[h]
\caption[figcal0]{The  Chern-Simons number as a function of
the time $t$ for the  gauge configurations \rf{cal5} with $f(t) = t/T$.}
\label{figcal0}
\end{figure}
\begin{figure}[h]
\caption[figcal1]{ (a): The
lowest eigenvalue as a function of time $t$ for the gauge configurations
\rf{cal5} (case (1)) $h_y= 0$, 0.5, 1.0 and 2.0. (b): The
chirality $\tilde{\Gamma}_5$ for the same values of $h_y= 0$ as in (a).}
\label{figcal1}
\end{figure}
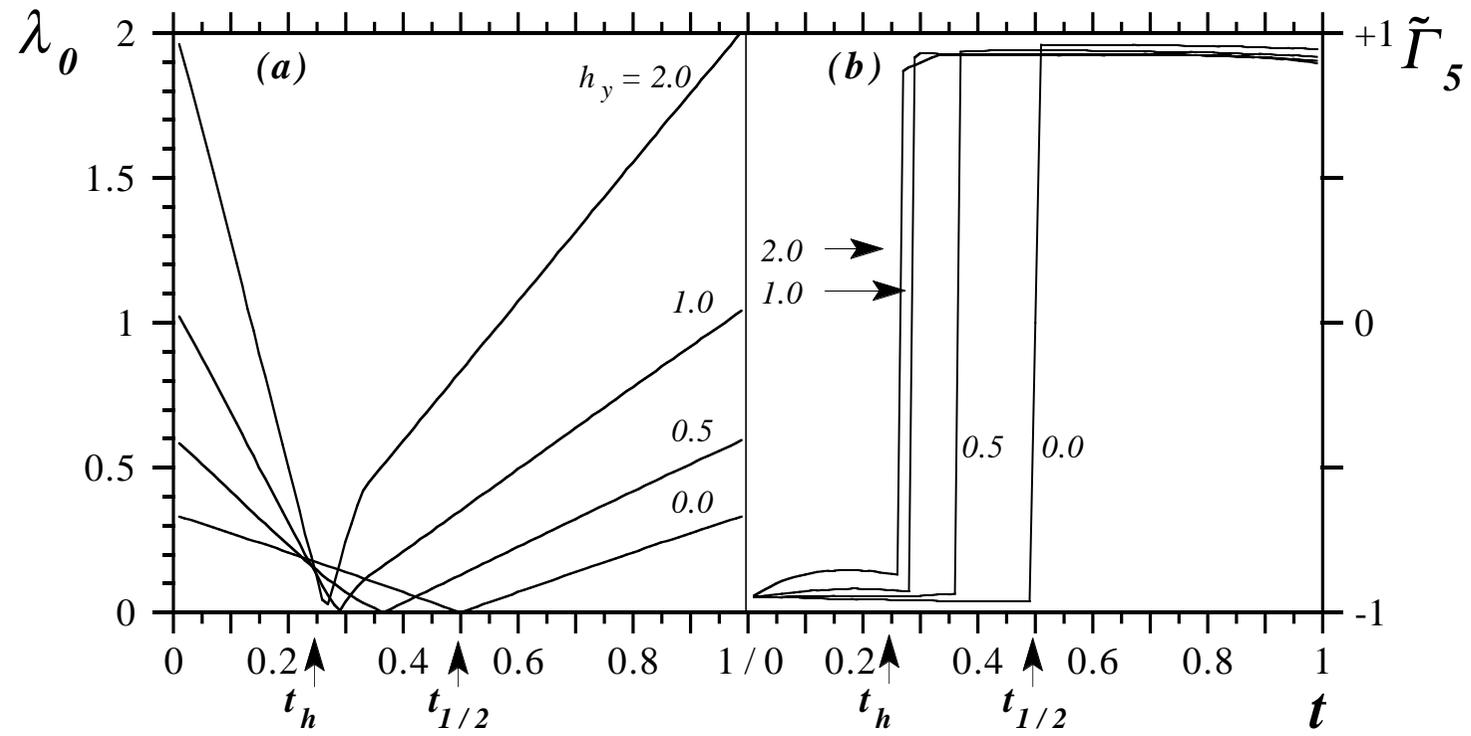
\begin{figure}[h]
\caption[figcal2]{(a): The lowest eigenvalue  in case (2)
for $h_y =0$, 0.5, 1.0 and 2.0 as a function of the time $t$.
(b): $\tGf$ for the same values of $h_y$.}
\label{figcal2}
\end{figure}
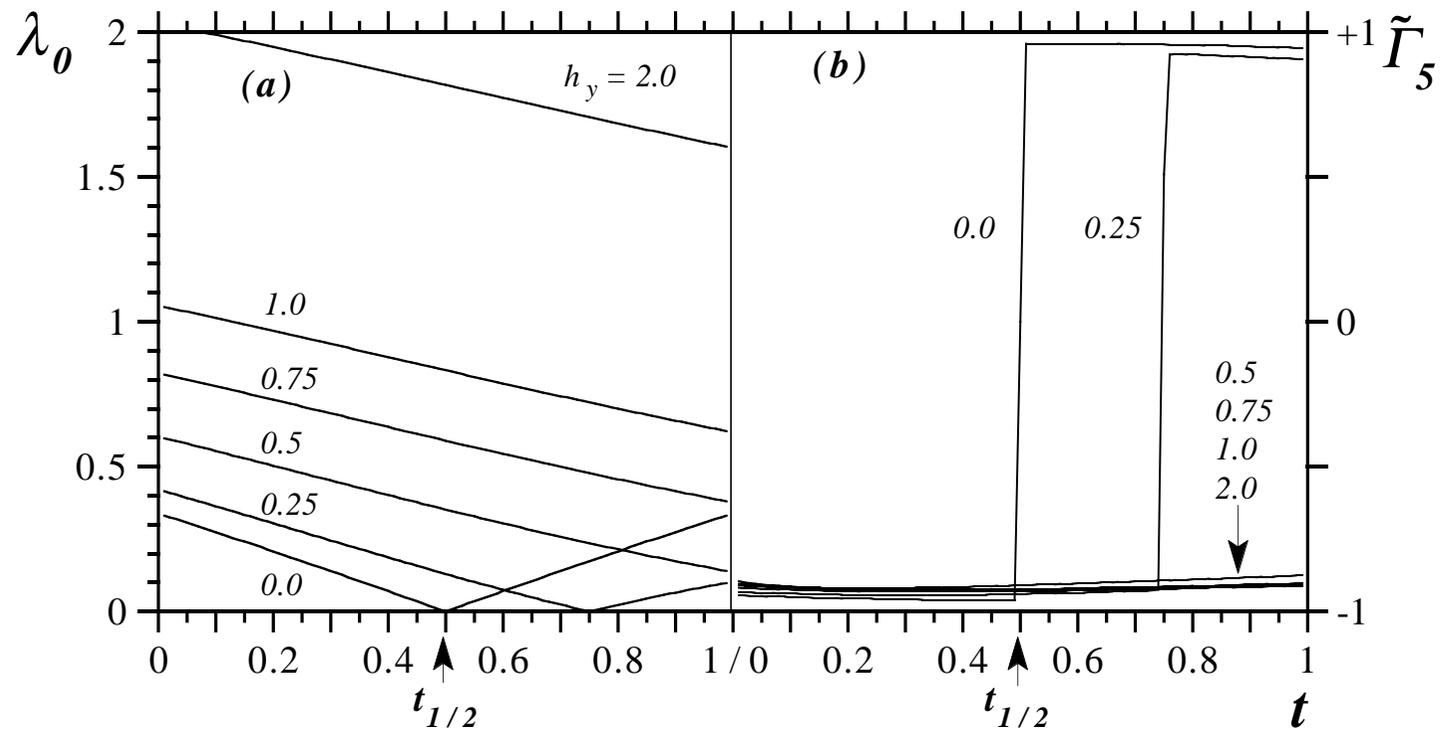
\begin{figure}[h]
\caption[figcal3]{(a): The lowest eigenvalue in case (3) for
$h_y=0$, 0.25, 0.5, 0.75, 1.0 and 2.0 as a function of the time $t$.
(b): $\tGf$ for the same values of $h_y$.}
\label{figcal3}
\end{figure}
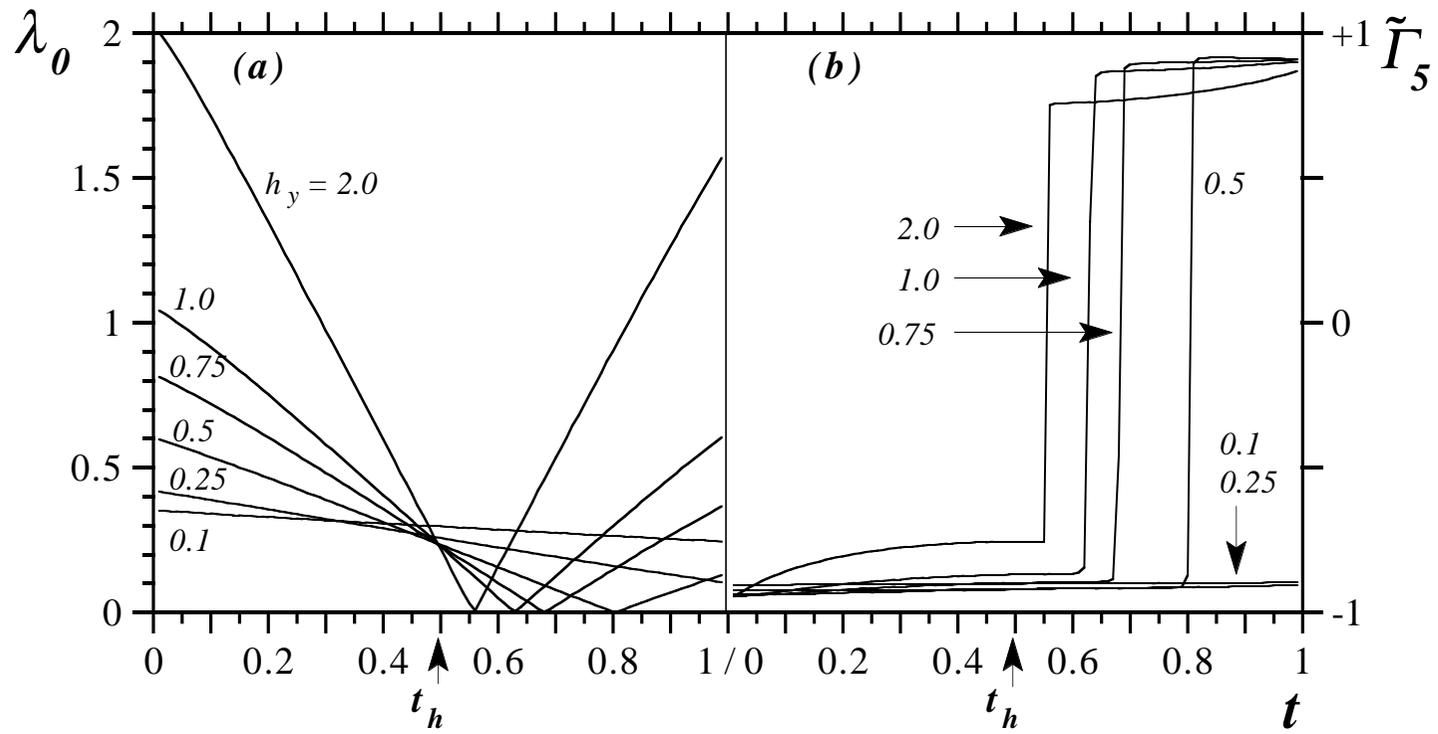
\begin{figure}[h]
\caption[figcal4]{(a): The lowest eigenvalue in case (4) for $h_y= 0.1$, 0.25,
0.50, 0.75, 1.0 and 2.0 as a function of time $t$. (b): $\tGf$ for the
same values of $h_y$. }
\label{figcal4}
\end{figure}
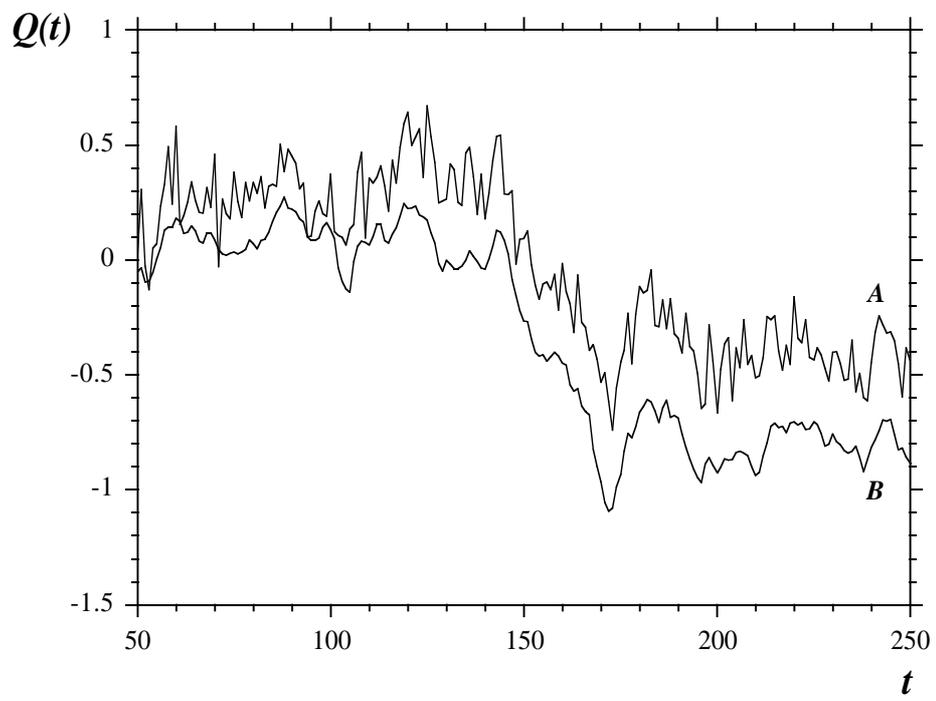
\begin{figure}[h]
\caption[fig1]{A typical ``sphaleron'' transition before relaxation and after
6 relaxation steps.}
\label{fig1}
\end{figure}
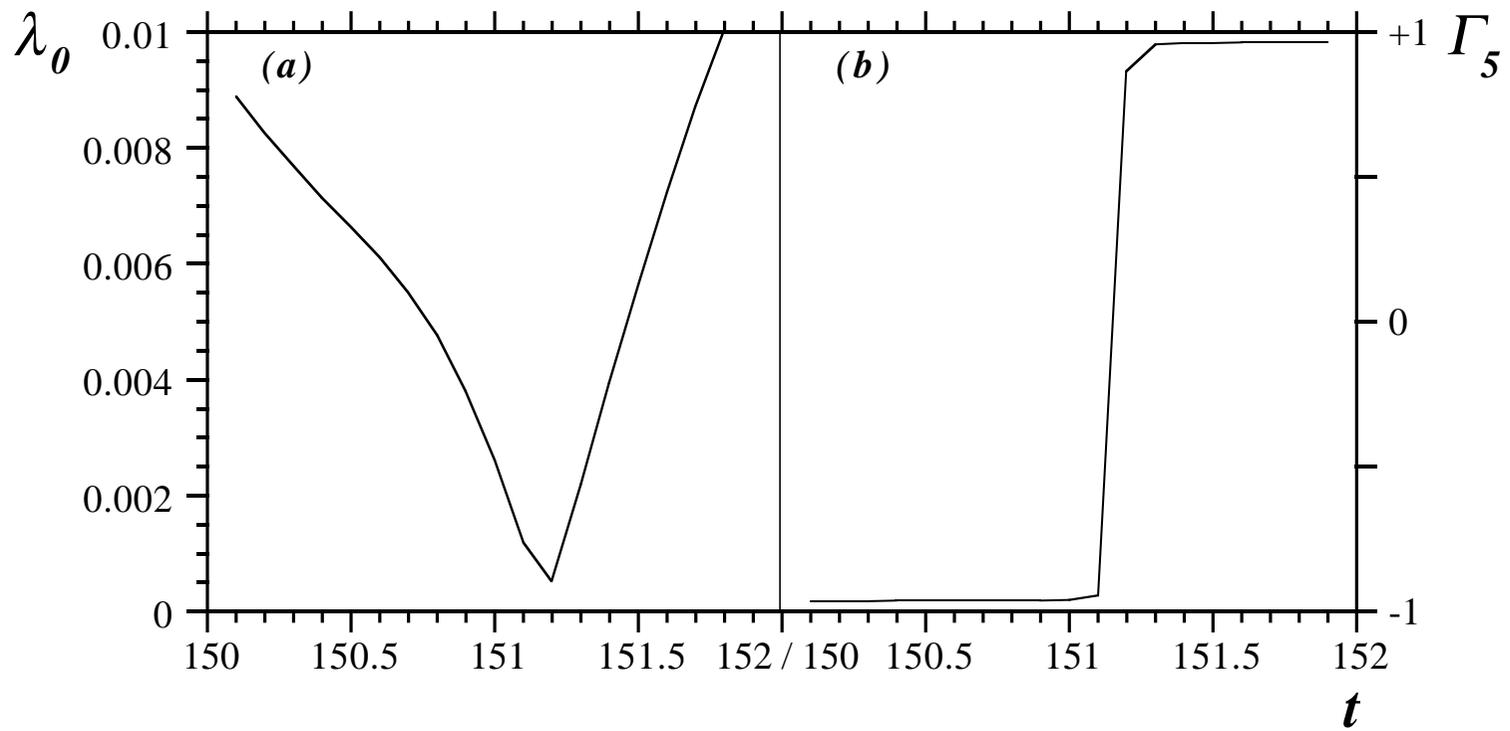
\begin{figure}[h]
\vspace{0.2cm}
\caption[fig2]{(a): The ``crossing'' of the lowest eigenvalue and (b):
the change in chirality for the transition shown in fig. 1.}
\label{fig2}
\end{figure}
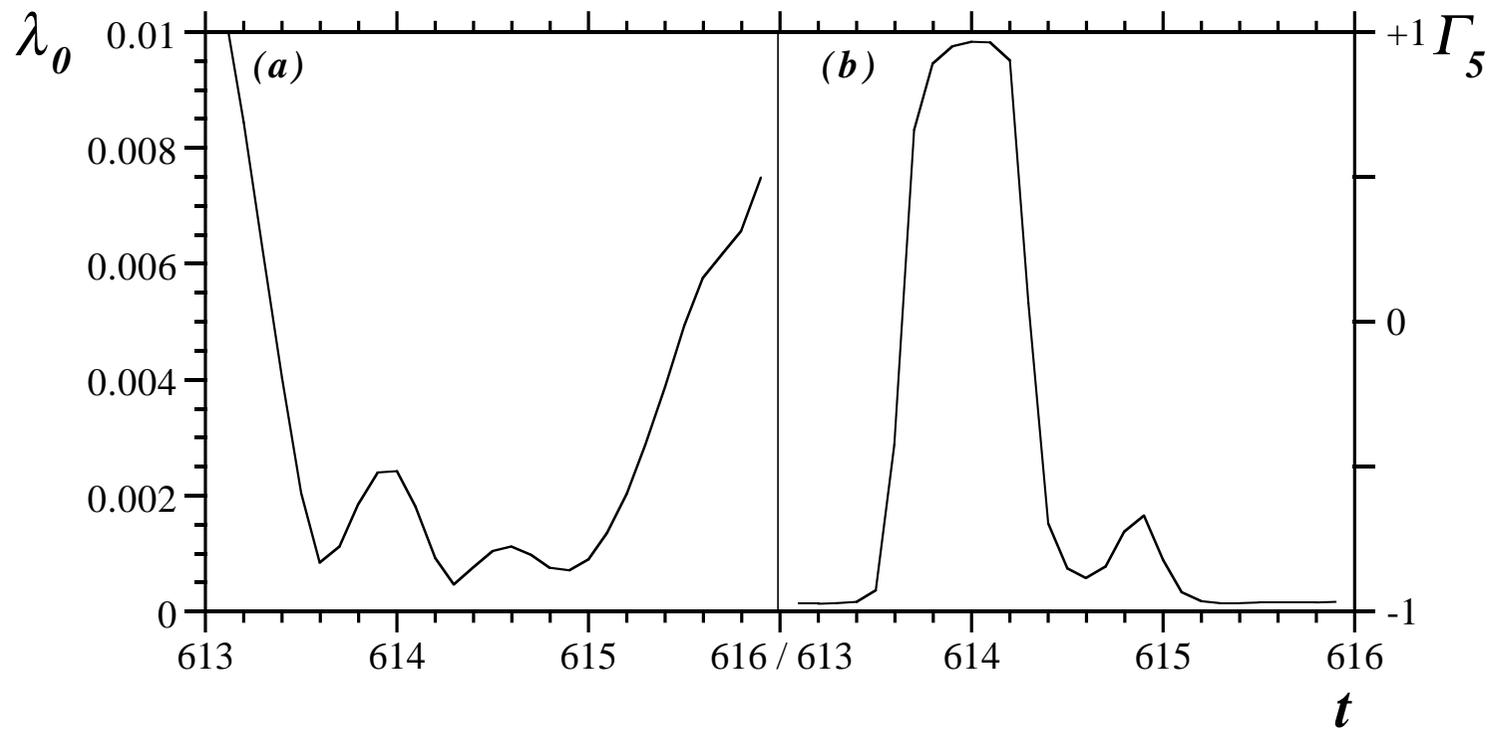
\begin{figure}[h]
\caption[fig3]{(a): The lowest eigenvalue for a sequence of
configurations which reach the ``sphaleron peak''
corresponding to $\Delta N_{CS} = 1/2$,
and the return to configurations with $\Delta N_{CS} = 0$. (b):
The change in chirality for the same configurations.}
\label{fig3}
\end{figure}
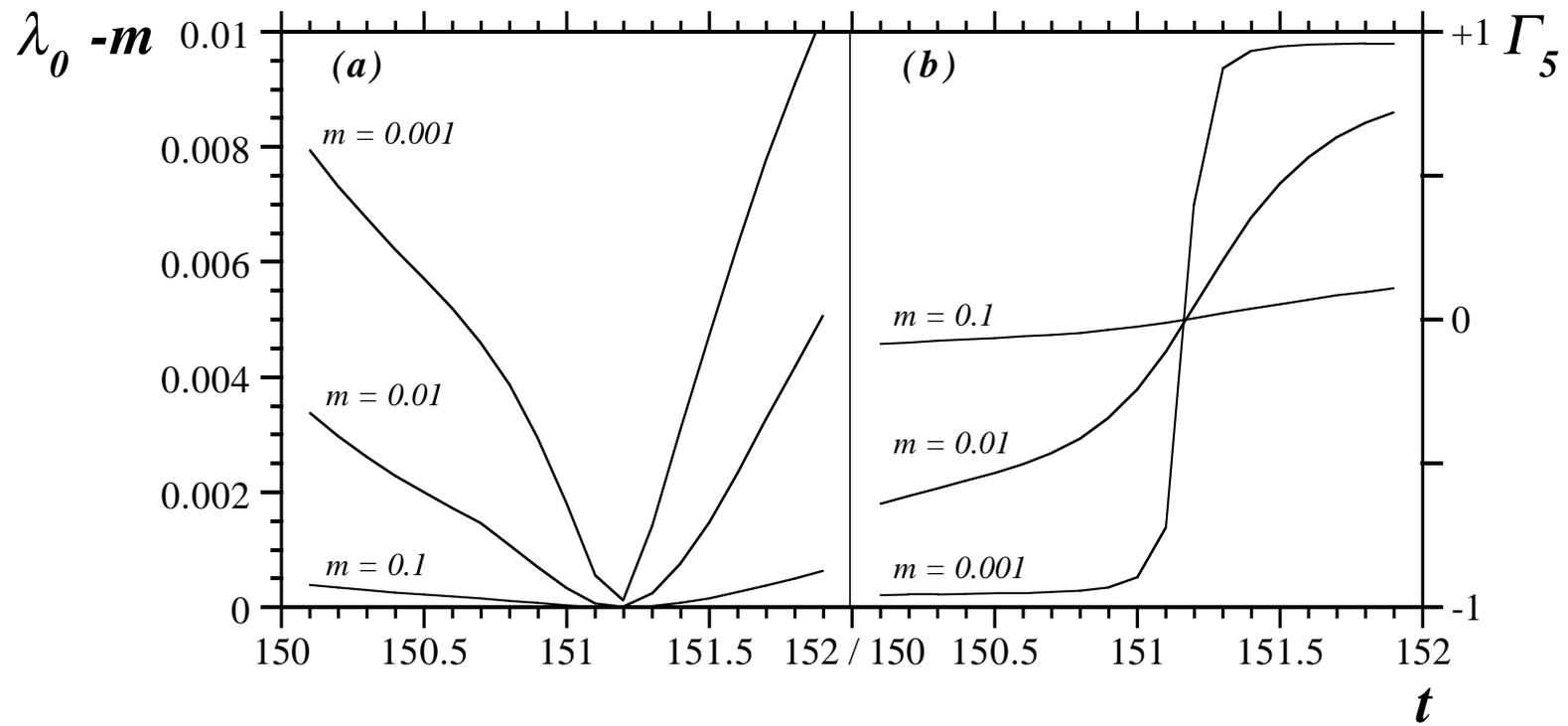
\begin{figure}[h]
\caption[fig4]{(a): The lowest eigenvalue $\l_0$ of
the massive Dirac equation \rf{*3.8}
for different choices of the mass close to a sphaleron-like transition.
On the figure we shown the difference $\l_0-m$ which is always positive.
(b): The chirality \rf{*mass} for the same configurations
and masses as in (a).}
\label{fig4}
\end{figure}
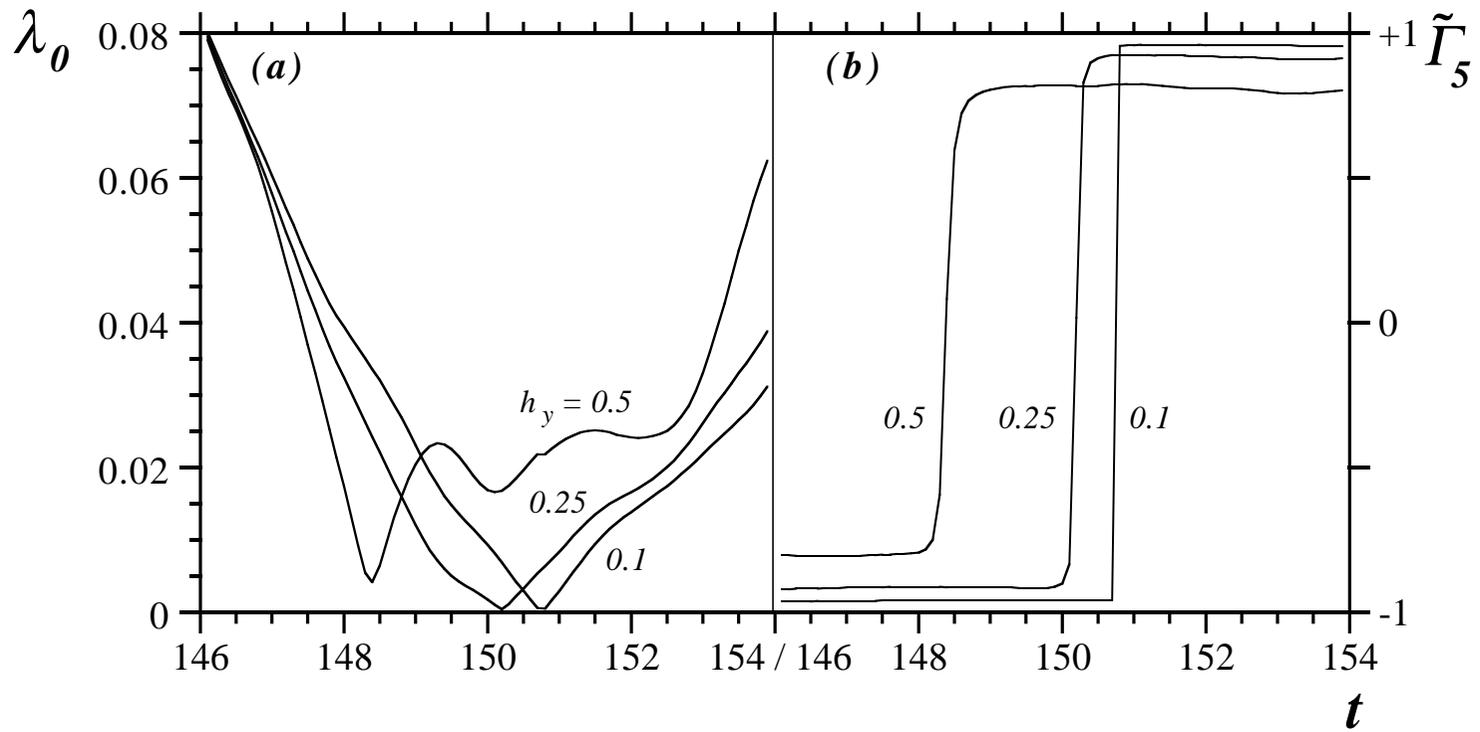
\begin{figure}[h]
\caption[fig5]{(a): The lowest eigenvalue near a sphaleron-like transition
for $h_y= 0.1$, 0.25 and 0.5.
(b): $\avGf$ for the same configurations and same
values of $h_y$ as in (a).}
\label{fig5}
\end{figure}
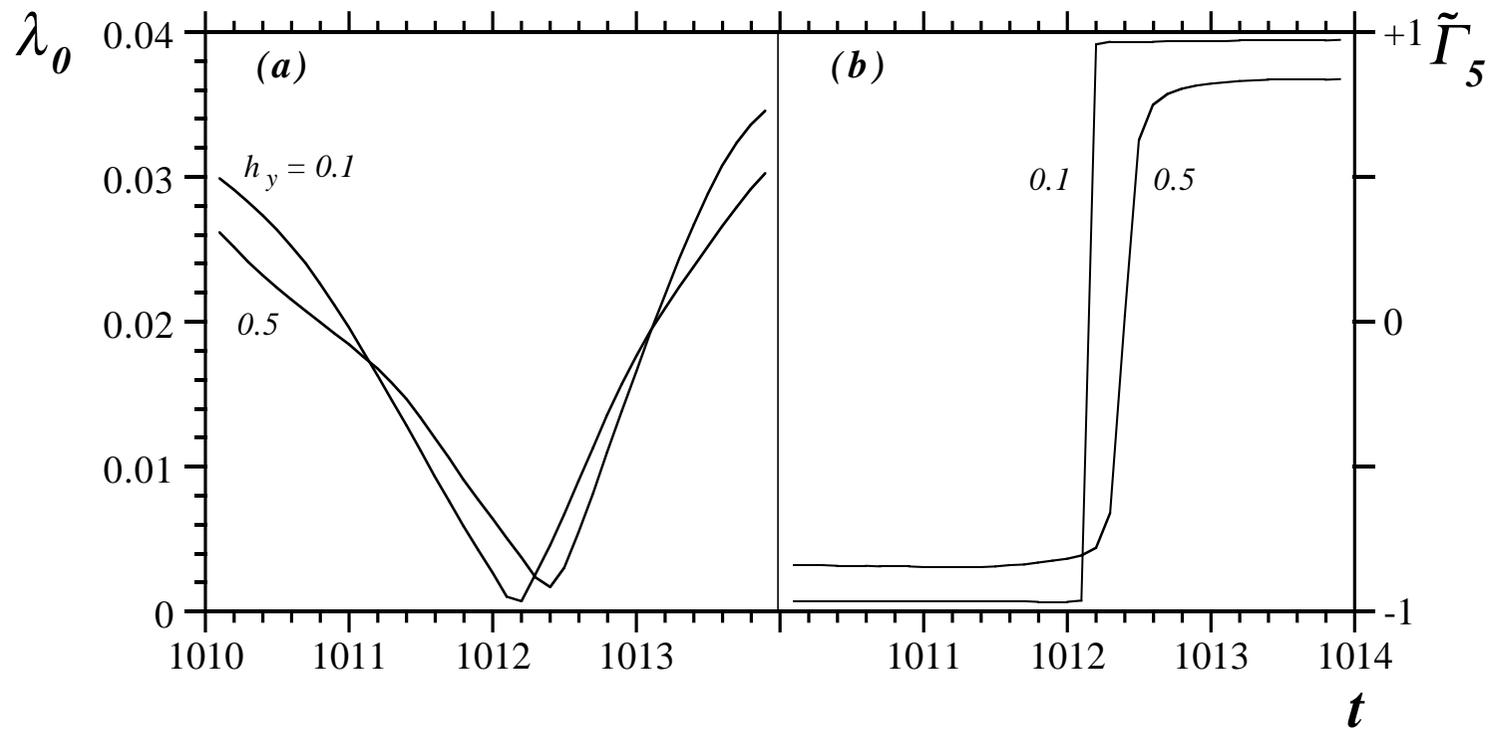
\begin{figure}[h]
\caption[fig6]{(a): The lowest eigenvalue for near a sphaleron-like transition
for $h_y=0.1$ and 0.5. The time $t_0$ where the eigenvalue crosses zero
has a much weaker dependence on $h_y$ than the transition considered in
fig. \ref{fig5}. (b): $\avGf$ for the same configurations and values of
$h_y$ as in (a).}
\label{fig6}
\end{figure}
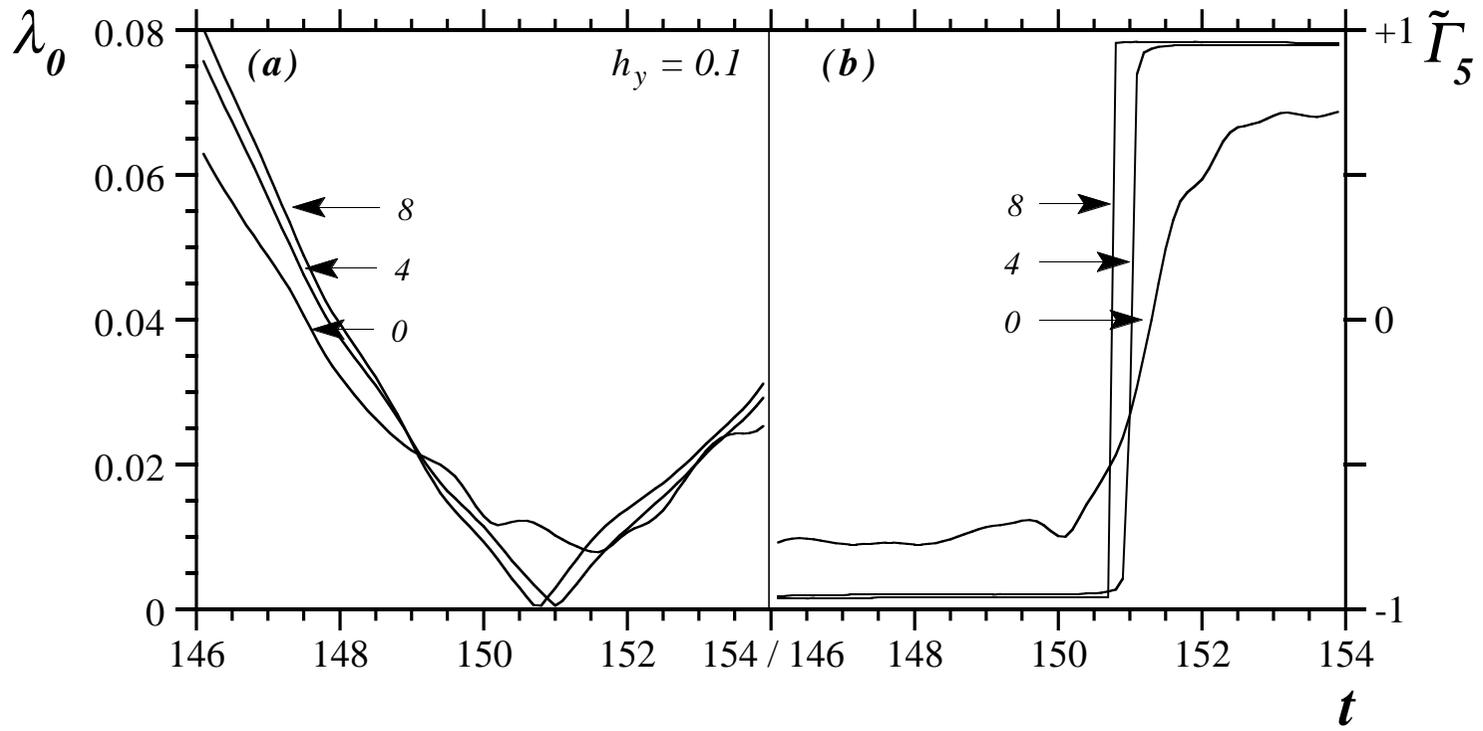
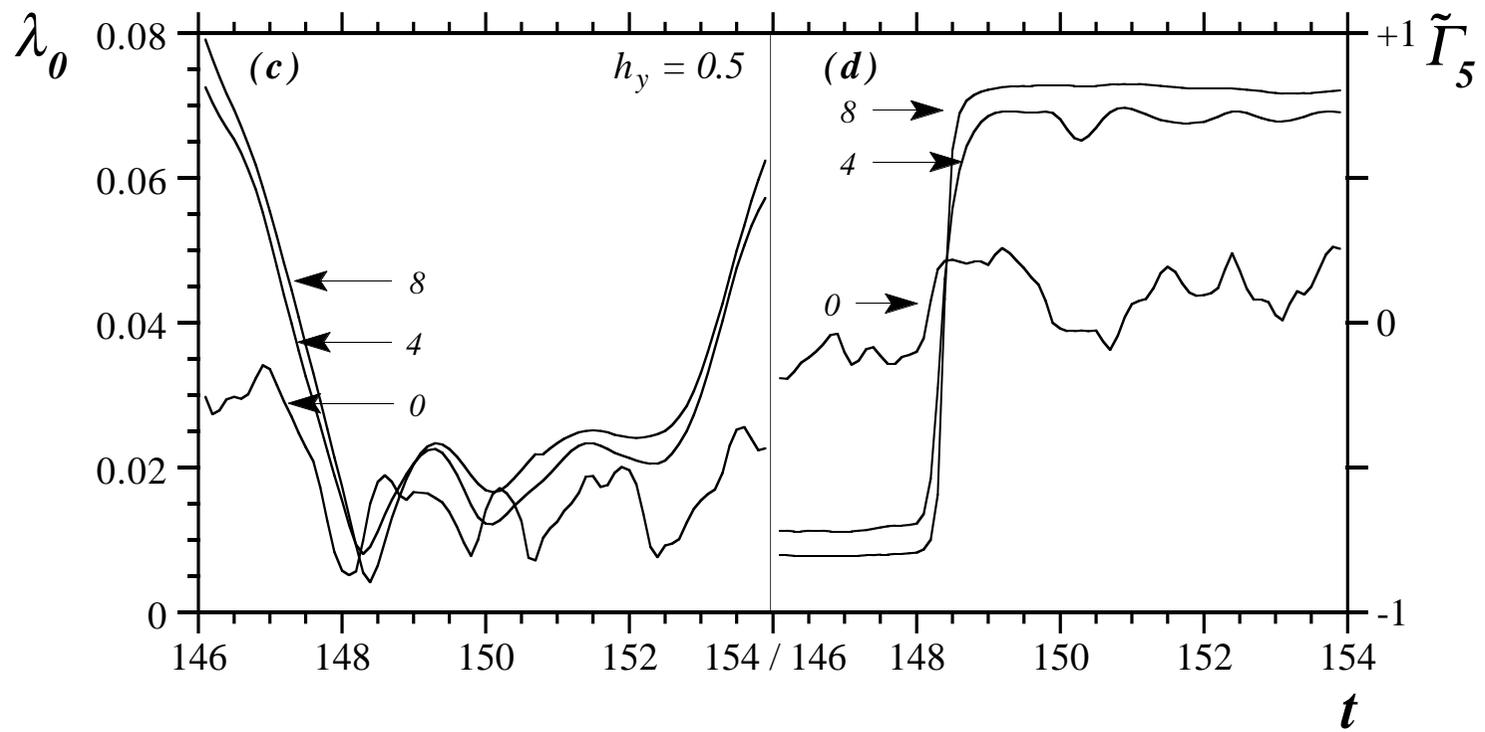
\begin{figure}[h]
\caption[fig7]{(a)-(b): The lowest eigenvalue and $\avGf$ for the
sphaleron-like transition used in fig. \ref{fig6} for $h_y=0.1$, if the
configurations has undergone 0, 4 and 8 relaxations. (c)-(d): The same
figures as in (a)-(b), only with $h_y=0.5$.}
\label{fig7}
\end{figure}

\end{document}